\documentclass[preprint]{aastex61}
\usepackage[latin1]{inputenc}
\usepackage{amsmath}
\usepackage{amsfonts}
\usepackage{amssymb}
\usepackage{amsthm}
\usepackage{gensymb}
\usepackage{graphicx}
\usepackage{color}
\usepackage[toc,page]{appendix}

\def\kms{~{\rm km~s^{-1}}}
\def\cmth{~{\rm cm^{-3}}}

\begin{document}

\title{A Fresh Look at Narrow-Angle Tail Radio Galaxy Dynamics, Evolution and Emissions}
\author{Brian J. O'Neill}
\affiliation{School of Physics and Astronomy, University of Minnesota, Minneapolis, MN, USA}

\author{T. W. Jones}
\affiliation{School of Physics and Astronomy, University of Minnesota, Minneapolis, MN, USA}

\author{Chris Nolting}
\affiliation{School of Physics and Astronomy, University of Minnesota, Minneapolis, MN, USA}

\author{P. J. Mendygral}
\affiliation{Cray Inc., Bloomington, MN, USA}

\begin{abstract}

We present a 3D magnetohydrodynamic (MHD) study of narrow-angle tail (NAT) radio galaxy (RG) dynamics, including passive cosmic ray electrons. We follow evolution of a bipolar-jet RG in a persistent crosswind through hundreds of Myr. We confirm traditional jet-bending models, while noting that our NAT exhibits a transitional morphology reminiscent of wide-angle radio  tails. Once deflected, jets remain internally stable, but are intermittently disrupted by external disturbances induced by the NAT dynamics itself. The disruptions enhance jet and tail magnetic fields. Disrupted jet plasma is heterogeneously mixed with denser wind plasma, yielding patchy, filamentary tails that grow longer at a rate exceeding the wind speed. Such fast tail extension could, for example, allow NAT tails to overtake extraneous ICM features, such as shocks and shear layers downwind of where the tails first form. Those events, in turn, could generate enhanced radio emissions within the ICM features themselves that do not follow the geometrical extension of the tails past the encounter. Analysis of synthetic radio observations reveals an extended time period once the NAT has developed in which it displays a nearly steady-state morphology with integrated fluxes that are roughly constant, along with a self-similar, curved integrated spectrum. In an appendix, we outline a simple analytic jet trajectory formalism with one adjustable parameter, using it to illustrate explicit trajectories that extend the classic bending model to arbitrary jet-wind orientations.

\end{abstract}

\section{Introduction} \label{intro}

Radio galaxies (RGs) are jet powered, ``radio mode'' manifestations of active galactic nuclei (AGNs) commonly found at both low and high redshift  in a variety of astrophysical  environments from voids to rich galaxy clusters. Although RG jets at their sources (super massive black holes in the centers of galaxies)  should  physically be bipolar and nearly axisymmetric, they frequently appear on kpc scales and beyond to be obviously distorted from this morphology \citep[e.g.,][]{garon19}. One of the common distorted forms shows them bent into so-called ``head-tail'' (HT) shapes with the AGN anchoring the head   \citep[e.g.,][]{burns79}. Often these HT sources appear roughly ``C'' or ''U'' shaped with extended, ``twin-tailed''  morphologies.  In case the twin tails asymptote to roughly parallel trajectories away from the head they are labeled ``narrow-angle tail'' (NAT) RGs \citep[][]{rudnick77}. Our purpose in this paper is to examine in some detail the dynamical elements that lead to the NAT morphology and to explore in simple terms potentially observable behaviors that might help to isolate the essential underlying physics.

NAT morphologies have long been modeled as the consequence of ram pressure from a crosswind, $\rho_w v_w^2$, producing a transverse pressure force that deflects the AGN jets ``downwind'' \citep[][]{brb79,jo79}. Here $\rho_w$ is the density of the wind, while $v_w$ is the transverse velocity of the wind relative to the AGN jet. The crosswind may represent motion of the AGN host galaxy through its external environment, ``weather'' motions within that environment, or a combination of the two. The detailed bending dynamics depends, for instance, on whether a wind interacts directly with the jet, or, perhaps, the jet responds to a strong ambient pressure gradient induced by a wind or some other large scale ambient structure \citep[e.g.,][]{jo79,gan17}. Our focus here is on direct cross-wind interactions. The inclination between the wind and the jet velocity also influences the jet trajectory (see the Appendix in this paper). However, a simple but effective cartoon of the bending comes from noting that the above ram pressure applied to an element of jet with longitudinal momentum density, $T_j$, jet radius, $r_j$ and differential length, $\delta \ell$ for a time interval, $\delta t$, transfers a transverse momentum to the jet element comparable to its original longitudinal momentum if $\delta t \sim r_j T_j/(\rho_w v_w^2)$. If the  jet velocity is $v_j$, the jet element being deflected will have propagated a length $\ell_b \sim \delta t v_j$ during $\delta t$. This leads to a characteristic bending length, $\ell_b \sim r_j ~(T_j v_j)/(\rho_w v_w^2)${\footnote{The Appendix outlines a quantitative treatment of jet bending, revealing that, while $\ell_b$ provides a reliable bending length standard, the shape and full extent of a bent jet structure depend on additional dynamical factors.}} \citep[e.g.,][]{jnom17}. Written this way, the expression for $\ell_b$ would apply to a supersonic, relativistic  mass dominated jet using $T_j = w_j \Gamma_j^2 v_j/c^2$, where $w_j$ is the (rest frame) jet enthalpy density, and $\Gamma_j$ is the Lorentz factor of the jet motion. For a Poynting-flux-dominated jet, $T_j = S_j/c^2$, where $S_j$ is the Poynting energy flux, while $v_j = c$. Then $\ell_b \sim r_j S_j/(c \rho_w v_w^2)$ \citep[][]{gan17}. In the case of a non-relativistic, supersonic mass-dominated jet, such as we simulate in the present work, $T_j = \rho_j v_j$, which then leads to the commonly cited bending length, $\ell_b \approx r_j~ (\rho_j v_j^2)/(\rho_w v_w^2)$ (cf. also the Appendix). 

We note, as well, that the above derivation makes obvious the somewhat counter-intuitive fact that all of the momentum in the deflected jets transverse to their original propagation direction is derived from the transverse wind, even though the wind velocity can be orders of magnitude smaller than the jet velocity. That is, the wind effectively transfers momentum to the jets, concentrating it in order to deflect their motions. We shall see below that once the jets have been deflected, there remains a complex momentum exchange in both directions between the jets and the wind that embeds them.

For a non-relativistic, supersonic jet, it is trivial from the above derivation to show,  using the standard Mach number relation $M^2 = v^2/c_s^2$, that when the jet and wind plasmas have the same adiabatic index, then $\ell_b \approx r_j (M_j^2  P_j)/(M_w^2 P_w)$, where $c_{s}$ is the sound speed. In this expression $M_j$ and $M_w$ are the internal jet and wind Mach numbers, while $P_j$ and $P_w$ are the respective pressures. If the jet is in pressure balance with the surrounding wind ($P_j = P_w$), it is thus evident that $\ell_b/r_j \sim (M_j/M_w)^2$. This last, simple scaling relation actually applies even if the jet is relativistic, provided one uses an appropriate, relativistic definition of the jet Mach number \citep[e.g.,][]{jnom17}. 

We point out that the only length scale needed to express wind-blown jet trajectories, so the physical scales of a RG formed in this way, is $\ell_b$, which scales with $r_j$ along with the Mach number ratio, $(M_j/M_w)^2$. Thus, for example, the physical size of the RG in our simulation described in \S \ref{SimDet} \& \S \ref{NATevo} could be scaled arbitrarily up or down for fixed ratio $\mathcal{M}_j/\mathcal{M}_w$, by resetting the jet radius, $r_j$. Alternatively, $\ell_b$, so the physical size and shape of the simulated RG, could be kept fixed while $r_j$ was rescaled, by varying $\mathcal{M}_j/\mathcal{M}_w \propto 1/\sqrt{r_j}$. One should keep in mind, however, that the associated dynamical time to reach a given stage in morphological development scales as $t_{dyn} = \ell_b/v_w$. So, if, for example, $\mathcal{M}_j/\mathcal{M}_w$ is kept fixed, while  $\ell_b$ is reduced by reducing $r_j$, then $t_{dyn}$ is reduced in proportion to $r_j$. As we shall see below, bent jets become episodically unsteady with timescales related not only to $t_{dyn}$, but also to shorter times connected to the jets. Consequently, time-dependent structural details can be complex. In addition, since synchrotron spectral aging effects depend also on time, resizing details of the radio spectral distributions to match a structural rescaling would require adjusting the aging timescales the same as $t_{dyn}$.

The validity of the above ``cross-wind-ram-pressure-bending'' cartoon has been confirmed in multiple simulation studies involving non-relativistic  plasma jets \citep[e.g.,][]{wg84,bal92,loken95,heinz06,porter09,mors13,jnom17}. Simulations of Poynting-flux-dominated ``magnetic tower" jets in dynamically active cluster media have shown analogous behaviors \citep[][]{gan17}. ``Aligned-wind''-based scenarios that lead to single-tailed  HT sources have also been discussed in \cite{jnom17} and \cite{nolt19} although those outcomes are outside the scope of the present paper.

While the basic picture of the ram pressure deformation model for twin tailed RGs is established, there is surprisingly little discussion in the literature exploring such important physics as the post-bending dynamical properties of the jets, the dynamical characteristics of the tails themselves, or the behaviors of magnetic fields and relativistic cosmic ray electrons (CRe) that are responsible for radio emissions used to establish the nature of observed sources presumed to represent such events. Nor has there been much discussion of the dynamical transitioning from relatively straight jets to the final HT morphology. Our purpose in this paper is to explore a number of these physical behaviors  in the specific context of NATs using high resolution (non-relativistic) MHD simulations incorporating energy-dependent transport of passive relativistic electrons (CRe), in order to provide a better basis for comparisons between observations and predicted model behaviors. Many of the same issues would apply qualitatively in some less-strongly-bent ``wide-angle tail" (WAT) sources. In fact, we see below that our reference simulated NAT actually passes through a dynamical stage where it would be difficult to distinguish it from a WAT source.

As an extension of the NAT RG formation context that is our focus in this paper we mention that multiple, deformed RGs have been observed in the vicinity of X-ray, cluster merger shocks whose appearances suggest pre-shock HT morphologies \citep[e.g.,][]{bona14,shim15}. At the same time, the detailed surface brightness  and spectral distributions  of some giant radio relics (elongated diffuse radio sources in cluster peripheries), as well as the relative infrequency of these radio relics compared to expected presence of merger shocks,  present an apparent theoretical problem if the relics represent CRe accelerated directly from ICM plasma by merger shocks \citep[e.g.,][]{vanw17}. The plausible association of RG with some relics has, in fact,  led to the suggestion that the relic radio emissions may be explained by populations of ``fossil" CRe that were released by the RG, cooled radiatively, and then reaccelerated by the passing shock \citep[e.g.,][]{shim15,kr16}. The highly elongated morphologies common to NATs make them obvious candidate RGs for such a scenario.

The present paper is part of a simulation study inspired by such possibilities as these. We analyze from simulations NAT formation in a steady crosswind. In a companion paper \citep[][]{on19b} this NAT then provides initial conditions for simulations of cluster merger shocks overrunning NATs, providing probes of the NAT-cluster-relic scenario. In additional, related work, \cite{nolt19} and \cite{nolt19b} study simulations of merger shocks  running over lobed RGs formed in initially stationary ICMs.

The remainder of this paper is organized as follows: \S 2 details the simulation setup. In \S 3, we analyze the dynamical evolution of our simulated NAT. In \S 4, we examine synthetic observations of our NAT. A summary of conclusions is presented in \S 5. We also include an appendix that outlines a simple analytic dynamical model for ram pressure jet bending for arbitrary alignment between a crosswind and the AGN jet axis and compares it to 3D simulation results.

\section{Simulation Details} \label{SimDet}

The work presented here is based on 3D MHD simulations carried out with the WOMBAT code described in \cite{mendthesis}. The code utilizes the Roe-based, 2nd order accurate total variation diminishing scheme for solving the ideal non-relativistic MHD equations (MHDTVD) described by \cite{rj95} along with the ``constrained transport" method of \cite{ryu98} to maintain a divergence-free  magnetic field. 

Before listing specific physical model parameters we remind readers that, as outlined in the introduction, dynamical lengths and times associated with morphological properties of our simulated flows can all be scaled relative to the bending length, $\ell_b$ and the dynamical time, $t_{dyn} = \ell_b/v_w$. In particular, $\ell_b$ can be made smaller or larger than the nominal value ($\approx 35$ kpc) that results from the parameters listed below. {\emph{As a reflection of this scalability, lengths and times given in figures, which are based on nominal parameter values to follow, are marked by $^*$}}. In principle, it is also possible to rescale radiative loss timescales through adjustments in redshift and magnetic field strength normalization, although those adjustments would be more complex.

The simulation discussed below, which we label our ``reference NAT simulation''  used a Cartesian, Eulerian grid with cells, $\Delta x = 0.5$ kpc on a domain  spanning $x,y,z =\pm 513, \pm 121.5, \pm 283.5$~kpc.
This simulation domain was filled with a uniform, adiabatic, $\gamma = 5/3$, and  unmagnetized ICM (``the wind'') with density,  $\rho_w = 5/3 \times 10^{-28} ~{\rm g} \cmth$ and pressure, $P_w = 10^{-12}$ dyne  $\rm{cm}^{-2}$, (sound speed, $c_{s,w} = \sqrt{\gamma P_w/\rho_w} = 1000\kms$). While a homogeneous, unmagnetized ICM is not very realistic, those attributes allow us to isolate cleanly behaviors explicitly related to the magnetized jets that form the NAT. The ICM wind was given an initial flow velocity, $\vec{v}_w$, along $\hat{x}$. The wind was maintained by continuous  boundary conditions at the $-x$ domain boundary.  For our reference run $v_w = 900 \kms$ (Mach number, $\mathcal{M}_w = v_w/c_{s,w} = 0.9$).

The simulation also included a light, bidirectional, steady and supersonic jet pair oriented along the $\hat{z}$ axis (so orthogonal to the ICM wind) initialized at the start of the simulation ($t = 0$) and maintained until it was switched off towards the end of the simulation at $t = 546.6$~Myr. We note that while this is a remarkably long period of continuous AGN activity, observations of some head-tail radio sources approaching or even exceeding 1~Mpc with no apparent gaps in the tails (e.g., \citealp{owen14,wilb18}), would likely require nominal AGN active lifetimes of several 100~Myr.  Jet power variations, or even brief gaps, provided they were more than several Myr from the time of observation,  possibly could go undetected in the radio structures. Visible presence of the jets in the head connecting to the AGN obviously requires current AGN activity. Detailed power cycle analysis, or exploration of specific circumstances that might allow especially prolonged activity are beyond the scope of this paper, but will be explored in future work.

The jet flow was created in a stationary launch cylinder 24 grid cells, so $12$ kpc long, aligned with $\hat{z}$ (orthogonal to $\vec{v}_w$) and centered at position $(-390, 0, 0$)~kpc. Plasma conditions were maintained inside the cylinder in pressure equilibrium with its surroundings ($P_j \approx P_w$) and emerging as collimated, supersonic flow from each end. The outer radius of the launch cylinder was 4.5~kpc, including a 3 cell, coaxial ``transition collar'' to the ambient conditions. Reflecting boundaries for magnetic field and velocity were applied to the periphery of the launch cylinder. 

The jet plasma (adiabatic index, $\gamma = 5/3$) in this simulation emerged with density and velocity, $\rho_j = 10^{-2}\rho_w$ and $v_j = 2.5\times 10^4$~km s$^{-1}$, giving an internal, emergent jet Mach number, $\mathcal{M}_j = 2.5$. Tests of these jets analogous to those discussed in the Appendix established that their dynamics matched jets with the same density and velocity having radii, $r_j = 4.5$~kpc, so we, henceforth, use that radius for the jets in this simulation; i.e., $r_j$ matches the full radius of the launching cylinder.  The resulting nominal jet bending length, $\ell_b \approx (2.5/0.9)^2  r_j \approx  7.7 r_j \approx 35$~kpc (matching the analytic trajectory discussed in the Appendix). For use below we also define $r_{jc}$ as the jet core radius excluding the transition collar. In this case $r_{jc} = 3$~kpc. The total kinetic power of each jet is a little less than $10^{43}$ erg/sec.

A toroidal magnetic field, $B(r) = B_0(r/r_{jc})$ (uniform poloidal electric current inside $r_{jc}$) was maintained inside the launch cylinder, which transitioned to the ambient field (which could include previously launched jet fields) though the transition collar (there was a balancing effective return current inside the transition collar). The peak jet magnetic field strength, $B_0 = 1~\mu$G, yielded a fiducial $\beta_{pj} = 8\pi P_j/B_0^2 = 25$. The magnetic field strength in the launched jets was somewhat arbitrary, although it was selected to be in a range that allowed the jet power to remain kinetically dominant and to lead to evolved magnetic fields in the NAT tails roughly similar to $\sim \mu$G ICM magnetic field strengths. The evolved fields in the NAT are mostly less than $B_0$, although where stretching of the magnetic field lines is strong, it can exceed this by factors of a few, as we discuss in \S \ref{MagDyn}. As we shall see, with the chosen $B_0$ radiative cooling of relativistic electrons in the NAT are dominated by inverse Compton scattering of the CMB. Several times larger values of $B_0$ could lead synchrotron losses to dominate. In these contexts, we emphasize that our aims in this study are to examine a well defined example of NAT formation in order to understand the physics it illuminates, rather than to explore a full range of possibilities.

The simulation incorporated a passively advected, mass weighted\ scalar field, $C_j$, with $C_j = 1$ maintained inside the launch cylinder, but initialized to $C_j = 0$ elsewhere. In cells developing a  mix of jet and ambient fluid $0 < C_j <1$. Thus, $C_j$ represents the jet mass fraction in a given computational cell at a given time. Heuristically, $C_j$ acts like a jet pigment, whose intensity decreases as it becomes mixed with non-jet fluid. With this in mind, we, henceforth sometimes refer to $C_j$ as ``jet color". $C_j$ was evolved on the Eulerian grid according to the equation, $\partial C_j/\partial t + \vec{v}\cdot\nabla C_j = D C_j/D t = 0$, where $D /D t = \vec{v}\cdot\nabla $ is the Lagrangian time derivative operator. That is, the Lagrangian time derivative of $C_j$ vanishes. This contrasts with, for example, the equation of mass conservation, which can be written in similar form as $D\ log{(\rho)} /D t = - \nabla\cdot (\vec{v})$, where $\rho$ is mass density, and the RHS expresses the velocity divergence. This reveals that in a mass conserving system, the ``comoving'',  Lagrangian  rate of density  change is simply measured by the fluid velocity divergence \citep[e.g.,][]{landl}.

Passive, relativistic, cosmic ray electrons (CRe) injected from the jet launch cylinder were transported in space and momentum ($p \approx E/c = \Gamma_e m_e c$) using the  ``coarse grained momentum volume" (CGMV) transport algorithm \citep{jk05} for the CRe  distribution function, $f(\vec{r}, p, t)$, in the range $5~\rm{MeV}\le \Gamma_em_ec^2= pc\le 85~\rm{GeV}$, $10 \la \Gamma_e \la 1.7\times 10^5$.  The CRe momentum (energy) range was spanned by 8 uniform bins in $\ln{p}$.  At launch, $f(p) \propto p^{-q_o}$, with $q_o = 4.5$.  CRe adiabatic energy gains and losses were tracked, along with synchrotron and inverse Compton (iC)  radiative losses off the CMB. CRe passing through shocks were subjected to standard test-particle diffusive shock acceleration (DSA), for which the equilibrium downstream electron distribution is a power law, $f\propto p^{-q_s}$, with slope  $q_s = 4 M_s^2/(M_s^2 - 1)$. Since the typical timescale for DSA in the energy range being tracked is $\sim$~yr  \citep[e.g.,][]{bj14}, while simulation time steps, $\Delta t \sim \Delta x/v_{max} > 10^4$~yr, DSA was assumed to occur instantaneously at shock crossings. Shock strengths were measured at run time using methods similar to those presented in \cite{min01}. The CRe spectrum in each momentum bin was set to $q = min(q_i,q_s)$, where $q_i$ is the distribution slope in that bin upstream of the shock. The CRe spectrum emergent from the jet launch cylinder matches that behind a shock of Mach number, $M_s = 3$. We note that except for a very small population needed to avoid numerical singularities in the CGMV algorithm, the ICM itself ($C_j = 0$)  did not contain CRe. This simplification helps facilitate isolation of physics related to the jets themselves.

In order to track CRe iC losses off the CMB, it was necessary to fix the redshift, $z$, of the simulated RG.  The iC and synchrotron losses are equal when the magnetic field is $B = B_{CMB} \approx 3.24 (1+z)^2 \mu$G. In our simulations $z = 0.2$, for which $B_{CMB} \approx  4.67 \mu$G. Consequently,  for the parameters outlined above iC losses  usually, although not always, exceeded synchrotron losses.

The Appendix is focused on the dynamics of jet deflection  by obliquely impacting winds.  While primarily intended to outline a simple analytic model for jet bending, it presents results from multiple test simulations very similar to the reference case, but using somewhat smaller computational domains appropriate for their shorter dynamical times. The jet launch cylinder orientation, while remaining in the $x-z$ plane, was tilted for those simulations with respect to the wind flow at various angles, $0 < \theta_{ji} < 90^{\degree}$. Additionally, ICM and jet parameters (e.g.,~densities and flow Mach numbers) were varied to confirm the general validity of analytic scaling relations and their dependence on $\ell_b$ derived in the Appendix.

\section{NAT Evolution Overview} \label{NATevo}

\subsection{Dynamic Stages Outline} \label{evol-outline}

 Our reference NAT RG is, by design, kept as simple as possible in order to help isolate core behaviors that will likely contribute in real-world situations, but can easily be obscured in complex settings.  In that context, we note that the RG is powered by jets that remain on and unchanged for a prolonged time period and mention again that the jets (which are magnetized)  penetrate an unmagnetized and homogeneous wind blowing orthogonally to the jet launching axis. 
 
 Even though we argued in \S \ref{SimDet} that truly constant jets are probably not essential to the picture outlined below, obviously, if AGN activity is highly intermittent, the history is made more complex and the dynamical phases can be intermingled and even interacting. Such issues are outside our present scope.
 For introduction to NAT evolution in more complex wind environments see, e.g.,~\cite{morsony13}, who simulated NAT formation in a dynamical galaxy group, or \cite{porter09}, who simulated NAT formation in a moderately turbulent, magnetized, obliquely propagating ICM wind. The ICM turbulence and magnetization in the simulation reported in \cite{porter09} led to less distinct transition between the RG head and its tails and to enhanced tail radio emissions compared to what we see in the reference simulation, here, as one would expect. A wind more strongly turbulent, but turbulent on similar scales in comparison to the \cite{porter09} study, would possibly modify the forms of the bent jets in the RG head if the bending length, $\ell_b$, and the bending plane varied significantly within the head of the NAT, while the tails could become fatter and potentially more luminous, if, for instance the pre-existing and magnetized turbulence dominated motions within the tails (especially if the wind contained a population of CRe). Very much larger scale and long lasting pre-formed ICM turbulent structures could also lead to evident ``external'' radio features linking to the NAT if the encounters were ``glancing'', so did not disrupt those ICM structures. These more complex issues will be addressed in a future paper, but we do mention here the obvious point that the presence or absence of such features could possibly provide constraints on ICM turbulence properties. 
   \begin{figure}[ht]
        \centering
        \includegraphics[width=\textwidth]{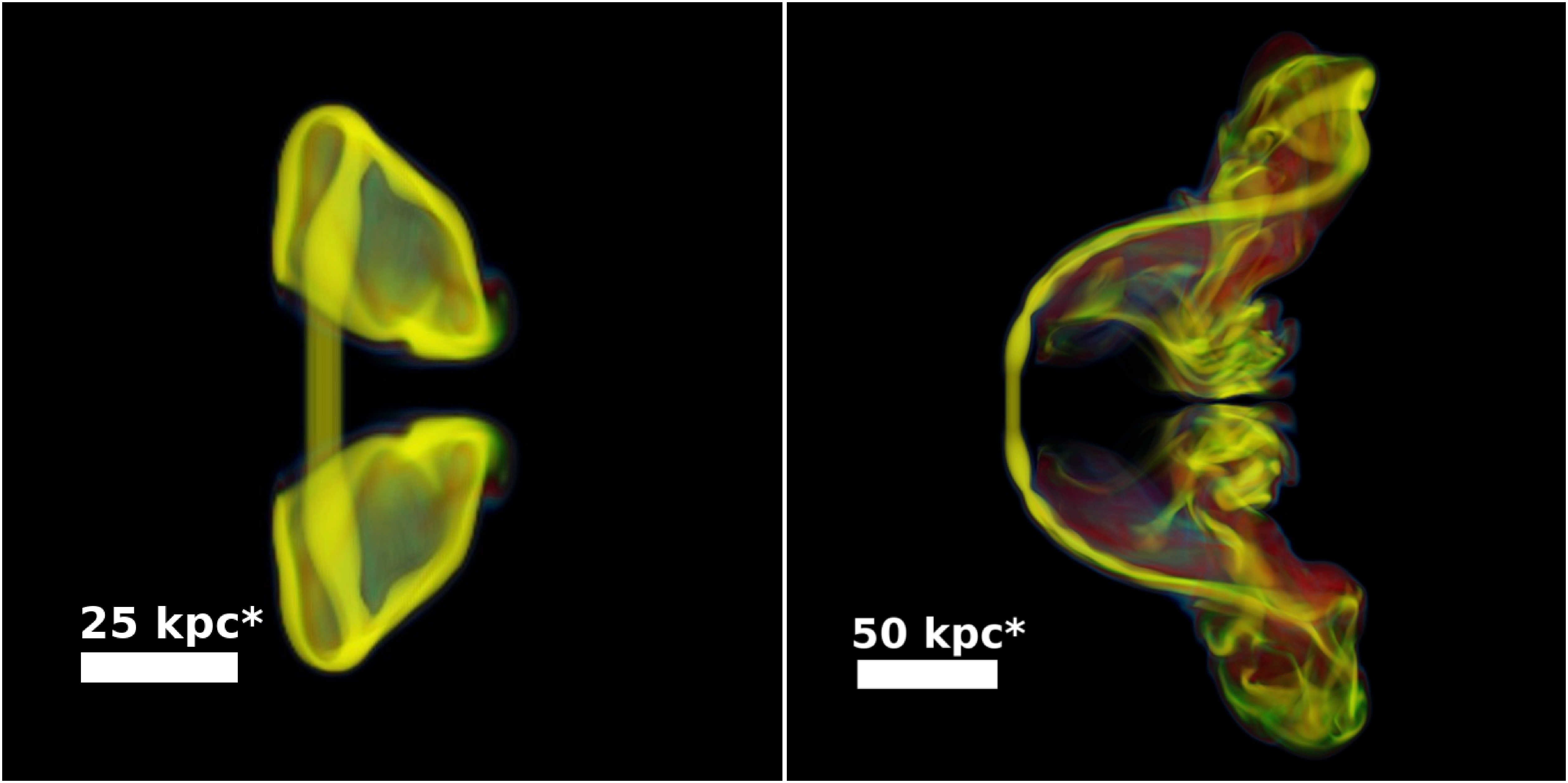}
        \caption{Volume rendering at early times of jet color, $C_j$, viewed along $\hat{y}$. Yellow identifies volume elements that are predominantly jet plasma followed by, in terms of decreasing mass fraction, green, red, and blue. Note that volume renderings are intended primarily for qualitative analysis, since multiple colors overlap along many lines of sight, and intensity reflects a line integral involving both color and opacity.  Left: Close up at $t = 16.4$~Myr$^*$. Distorted jet terminus plumes are evident. Right: $t = 98.4$~Myr$^*$. The transition towards NAT morphology is clear, along with the plume structures bridging between the tails and remnants of the initial jet flapping event. The $^*$ in labels serves as a reminder that the nominal length (and time) scales can be increased or decreased by adjustment of the bending length, $\ell_b$.}
        \label{earlycolor}
    \end{figure}

The history of our steady jets interacting with a homogeneous crosswind to form a NAT can be divided into two distinct phases while the jets are active, followed by a third, post-activity, decay phase.  The initial, first stage evolution begins as the light, but fast, AGN jets penetrate into the dense ICM and undergo deflection downwind by the crosswind flow as outlined in \S \ref{intro} and in somewhat more general terms in the Appendix (see also figures \ref{earlycolor}, \ref{earlyvslice}), or potentially even within the host galaxy ISM embedded in the same wind \citep[][]{jo79}. The ICM plasma in the downwind wakes of the transverse jets quickly becomes turbulent, even when the incident wind is not. By the time the jets have extended to lengths comparable to the nominal bending length, $\ell_b$, they show preliminary signs of transition towards NAT morphology, although that transformation turns out to be rather prolonged and complex.

    \begin{figure}[ht]
        \includegraphics[width=\textwidth]{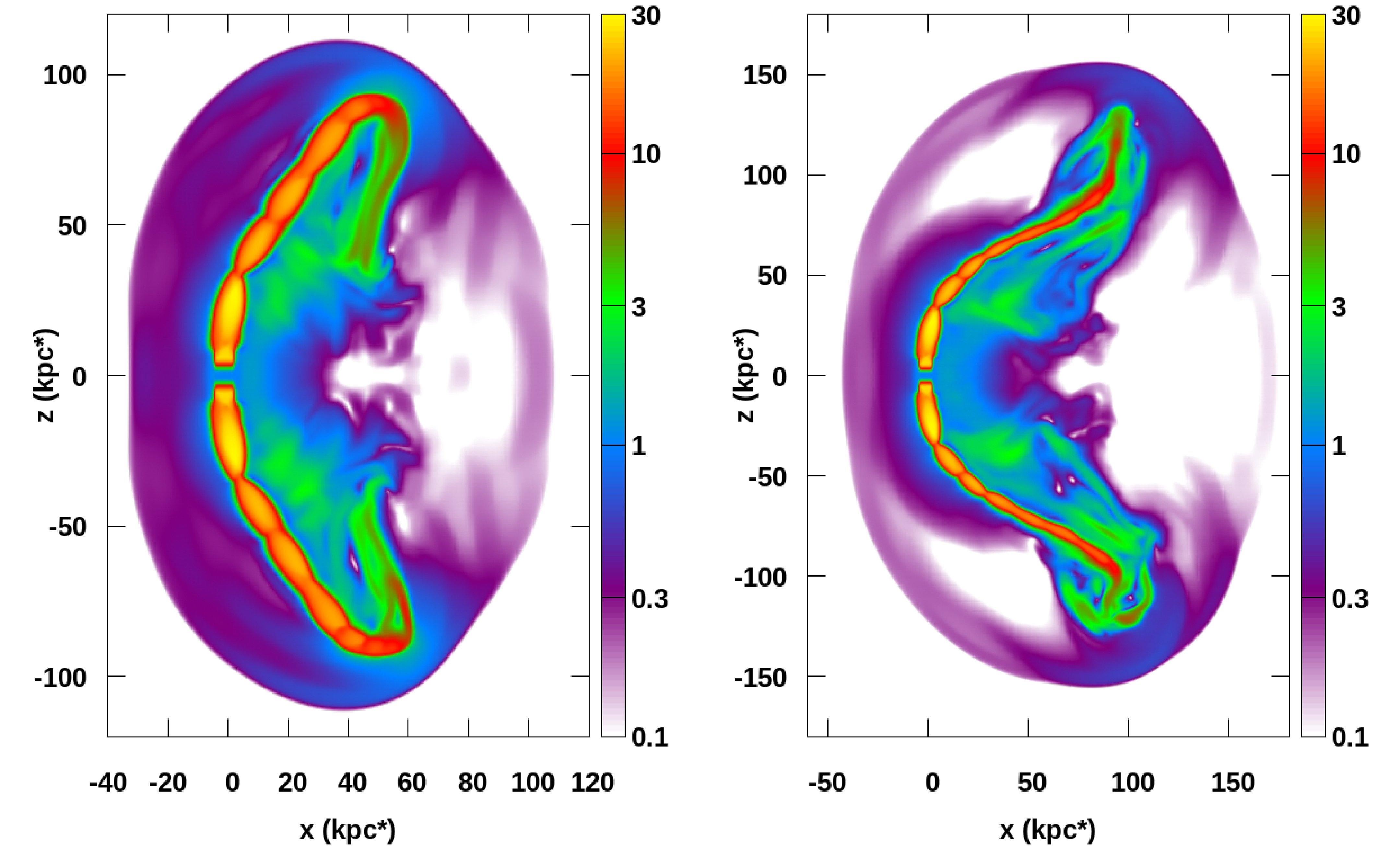}
        \centering
        \caption{Flow velocity (in units of $v_w$) in the $x$-$z$ plane at y = 0 measured in the undisturbed wind rest frame during the so-called transition phase. Left: $t = 49.2$~Myr$^*$. Right: $t =82.0$~Myr$^*$. The $^*$ in labels serves as a reminder that the nominal length (and time) scales can be increased or decreased by adjustment of the bending length, $\ell_b$. }
         \label{earlyvslice}
    \end{figure}

The clearest initial signs of the transformation come before the jet trajectories themselves begin to resemble their eventual arcs. Specifically, as illustrated in figure \ref{earlycolor}, jet plasma shed from each jet terminus is blown downwind, producing plume structures that eventually develop into distinctive and persistent NAT components\footnote{Note that volume rendering is designed to identify significant structures. It is generally not intended for quantitative analysis of intrinsic measures, since it involves a line integral, typically with total optical depth less than unity, so that, unless a single value is isolated, multiple intrinsic values typically blend along lines of sight.}. As the jets continue to extend, they actually undershoot their nominal equilibrium trajectory (they are less deflected than in their equilibrium trajectories). Subsequently, they abruptly ``overcorrect'' towards their equilibrium paths and are mostly pinched off (disrupted). This initiates long term, episodic ``flapping'' of the downwind jets that can become very large amplitude (figures \ref{earlyvslice}, \ref{wagfig}). When one of the jets is initially aimed upwind coming from its source, this trajectory over-correction-disruption event can be especially dramatic (see the Appendix). Even in our reference simulation with jets launched orthogonally to the wind the consequence of the undershoot just described leads to transient transverse extents approaching 150 kpc ($t \sim 80$~Myr), compared to the nominal bending length, $\ell_b \approx 35$~kpc (see figure \ref{earlyvslice}). The morphology at that stage is effectively that of a WAT. Subsequent jet disruptions after $t \ga 100$~Myr (see figure \ref{wagfig}) isolate these extensions as the jets adjust towards the long-term NAT morphology.  The remnants of these transition structures develop into distinctive vortex plumes at the end of each tail (see figure \ref{t045_obs}). As we see in \S \ref{sec:obs}, however, their radio emissions fade over time.

Over longer times the jet trajectories do average close to the simple analytic, equilibrium paths outlined in the Appendix; i.e.,~expected NAT morphologies develop.  However, jet flapping/disruption events continue to punctuate the dynamical evolution of the tails so long as the jets remain active. Over time both the tails and embedded jets extend farther downwind (see figure \ref{wagfig}). 

As a preliminary insight into the long term tail dynamics we make the obvious but significant point here that for jets initially orthogonal to the wind (as in our reference simulation) all of the momentum in the downwind-directed jet has been extracted from the wind. More generally, interaction between the jets and the wind in the head region transfers and concentrates wind momentum into the fast-moving jet plasma, which then transports that momentum downwind. The ``stolen'' momentum is eventually re-deposited in the tails downwind from where it is first extracted. Although on average the net momentum transfer should balance, as we shall see in \S \ref{sec:tails}, the plasma density in the tails is generally less than in the undisturbed wind. Consequently, the NAT tail plasma gets a net velocity boost out of the exchange. That is, the NAT tails are not passively advected in the wind, but actually propagate downwind faster than the wind. Our simulated scenario in this paper involves only a steady wind. {\emph{But, if, as in the scenario simulated by \cite{nolt19b}, the wind generating the NAT is a post shock wind, a consequence of the NAT velocity boost can be that the NAT tails actually overtake the shock.}} This, in fact, happens in the ICM-shock-generated NAT discussed in \cite{nolt19b}. That, in turn could enable NAT magnetic fields and CRe to become engaged in physics associated with the ICM shock transition (e.g., diffusive shock acceleration).
 
 Once the NAT morphology is fully developed, its structure can be spatially divided into distinct, dynamical regions. Proceeding from the AGN source into the tails these are:{\bf{~(1)}} the U-shaped coherent jet structure defining the head, {\bf{(2)}} a low pressure, low density  turbulent wake inside the ``U'', {\bf{(3)}} a dynamical transition region connecting the head region jets to the extending tails, which we dub the ``disruption zone", and within which the jets continue to be episodically disrupted, then reformed as they extend, and finally, {\bf{(4)}} the  extending, heterogeneous and turbulent tails, which also connect to prominent plumes developed relatively early in the source evolution and discussed in the next section. Although those plumes form out of the transition towards NAT behavior, they persist in association with the tails until the end of our simulation. However, their radio emissions fade over time as the CRe within them age.

\subsection{Jet Dynamics} \label{JetDyn}
\subsubsection{Early Evolution}
\label{sec:early}

\begin{figure}[ht]
\centering
\includegraphics[width=\textwidth]{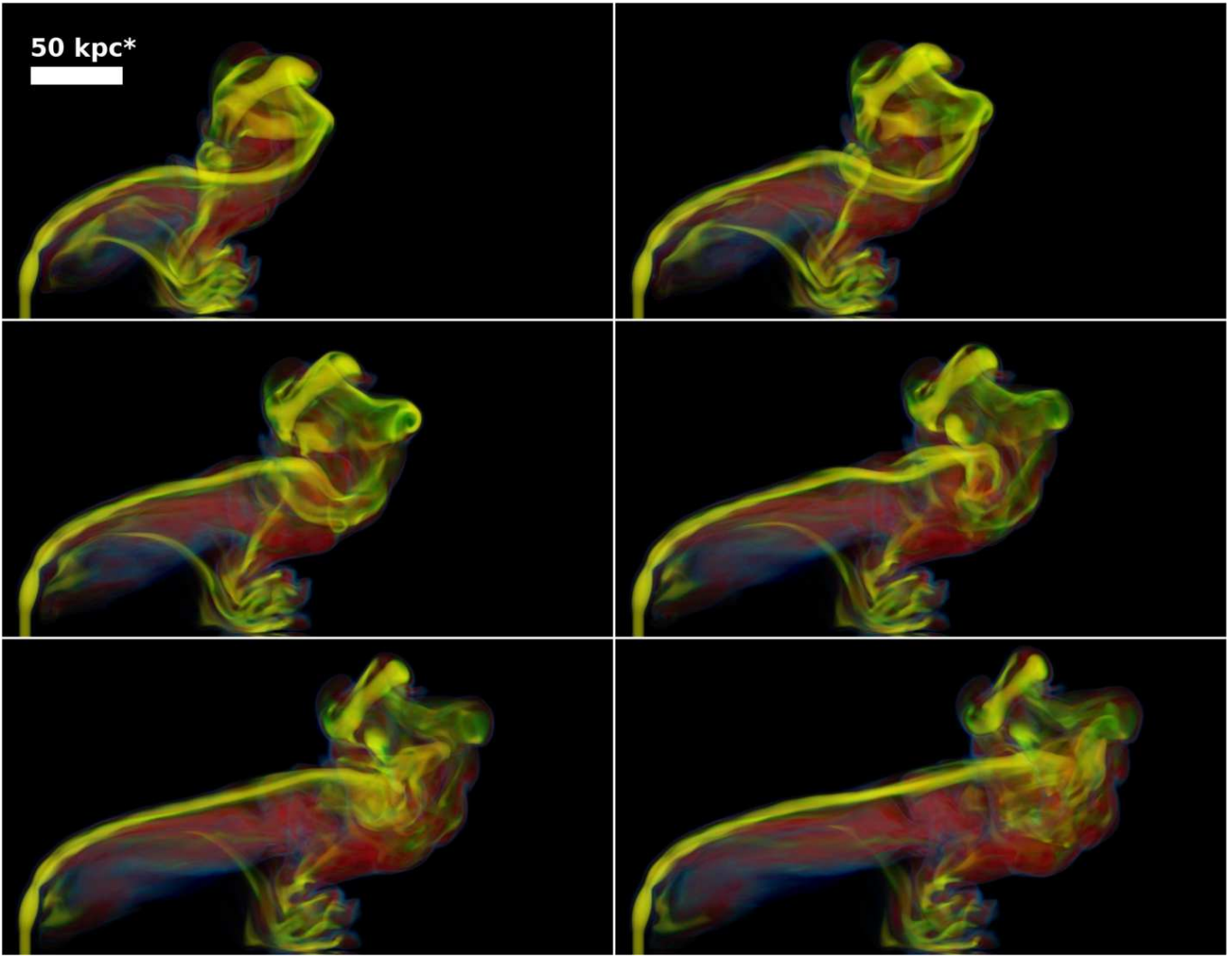}
\caption{Volume renderings of jet color, $C_j$, demonstrating the sequence of a disruption event. The upper left image is at $t = 114.8$ Myr$^*$, with successive images separated by 16.4 Myr$^*$. So, the bottom right image corresponds to $t = 196.8$ Myr$^*$. Color scale is the same as figure \ref{earlycolor}. The $^*$ in labels serves as a reminder that the nominal length (and time) scales can be increased or decreased by adjustment of the bending length, $\ell_b$.}
\label{wagfig}
\end{figure}

In simple terms a supersonic, AGN  plasma jet behaves as a low density ``beam'' driving into the denser background ICM. In the case of a static ambient ICM, the jet plasma, after passing through a jet termination shock, would flow back around the propagating jet, displacing the ICM to form a classic, low density radio lobe. In the NAT context, with a strong ICM crosswind, the plasma emerging from the jet terminus is unable to form such a lobe. Initially, when the jets still propagate roughly straight and  the terminal flow is across the incident wind, highly distorted ``mushroom cap" terminal plasma plumes form{\footnote{See \cite{nolt19} for the analogous analysis of jet propagation into a quasi-aligned ICM wind}} as shown in figure \ref{earlycolor}. Upwind portions of these terminal plumes are pressed against the jet. Some of the expelled plasma is entrained again in jet flow. Some of that becomes trapped just downwind, near the head of the source so long as the jets are active. The rest becomes incorporated into prominent, persistent vortex features creating a bridge-like structure between the two tails as NAT morphology develops (figures \ref{earlycolor}, \ref{wagfig}). We will discuss in \S \ref{sec:obs} synthetic radio observations of this structure. In short, it is relatively radio luminous as it first forms, but, because it becomes isolated from jet-provided fresh CRe, synchrotron emissions fade over time.

Propagation of the launched jets also generates a pair of merging shocks in the ICM, just as for a RG forming in a static ICM. In a static environment the combined, bounding ``RG cocoon shock'' would be roughly axisymmetric. That symmetry is broken in the NAT context, of course. Either way, the bow shock Mach numbers just beyond the jet termini roughly match the Mach number of the extension rate of the jet termini with respect to the local ambient medium. This depends, in turn, on the internal Mach number of the jet and the local ambient flow properties \citep[e.g.,][]{nolt19}. On the other hand, the lateral portions of these shocks are generally weak, with Mach numbers only slightly more than unity \citep[e.g.,][]{jnom17}. In the NAT reference simulation under examination here the wind velocity relative to the AGN is only slightly subsonic, $\mathcal{M}_w = 0.9$. Consequently, even before the jets are strongly bent downstream by the wind, the weak lateral shock structure barely propagates upwind from the AGN (the jet launch cylinder in the simulation), while it propagates downwind at roughly twice the wind speed in the AGN frame.  Once the jet terminations are deflected downstream, the bow shock Mach number is substantially reduced by geometry alone. Jet-associated ICM shocks become largely indiscernible from other features in the ICM flow.
    
There also is an ICM bow wave upwind of the jet launch cylinder and the transverse jet portions. In the reference simulation, where $v_w$ is subsonic, this bow wave never becomes a shock, so does not establish a stationary standoff location. Immediately upwind of the launch cylinder the ICM compression remains $\sim 50$\%. In a more realistic setting, the ISM of the host galaxy might provide an obstacle roughly comparable to the launch cylinder in our simulation.
    
\subsubsection{Transition of the Jets Towards a NAT Configuration}
    
As outlined at the beginning of \S \ref{evol-outline} the jets initially extend with somewhat less bending than predicted by the simple models utilizing an equilibrium bending length, $\ell_b$. But, they then adjust towards that trajectory, becoming unsteady, with episodic ``flapping''. As already noted, the scale and amplitude of jet flapping can substantially exceed the nominal bending scale, $\ell_b$. In the case of our reference simulation object, this transition phase includes a period when the RG morphology is essentially that of a WAT. As also noted above, the abrupt trajectory corrections associated with flapping are associated with jet disruption events; the jets become pinched where they bend sharply. The momentum within the pinched-off jet segments propagates inwards, towards the other jet flow region. On a timescale comparable to, or a bit less than the ICM sound crossing time between the two forming tails, the disrupted flow impacts the other jet, initiating a subsequent disruption event for that jet. In our reference simulation that time interval is $\sim 50$ Myr, which then roughly measures the interval between ``flapping events'' of each jet. In the reference simulation, where the two jet trajectories are qualitatively similar, the jet flapping/disruption events occur at very roughly the same times for the two jets. On the other hand, for obliquely launched jets associated with the NAT formations outlined in the Appendix, the two jet behaviors can be quite different. We mention this here especially in the context of transition towards the NAT morphology, although this behavior continues into the subsequent NAT phase of the source evolution. 

At this point also we mention in passing that the transverse scale of the RG during this transition would scale nominally with $\ell_b$, while the time interval within which the transition to something more obviously NAT should scale {\it{very}} roughly with $\ell_b/v_w$. However, since the transition depends in complex ways on both the dynamics of the individual jets as they are first becoming bent, as well as on their interactions with each other, it is difficult to write down reliable estimates of what that most likely length or the the transition time should be. In a more realistic external environment, other factors related to external, ICM structure likely would play significant roles, as well.

\subsubsection{NAT Phase Jet Dynamics}
\label{sec:natjet}

The above transition behaviors do ultimately lead to NAT morphologies in our simulations. We emphasize again that the episodic jet flapping and disruption events continue in the NAT phase. In fact, the disruption events are instrumental in defining the turbulent behaviors that develop in the tails. Roughly speaking the resulting outer turbulence scale in the tails is $\ell_t \sim \ell_b$  so in the reference simulation $\ell_t \sim 30$ kpc. The associated turbulent velocity is $\delta v \sim v_w$, giving a characteristic eddy time $\tau_{eddy} \sim \ell_t/v_w \sim (r_j/v_w) (M_j/M_w)^2$. In the reference simulation, $\tau_{eddy} \sim 35$ Myr. We mention in passing that turbulence in the wake region immediately downwind of the jets in the head region is driven on smaller scales closer to the jet diameter, and, although it does contribute to unsteady jet behaviors, does not appear to be their dominant driver.

There are two additional features of long term jet propagation worthy of mention here. First, our simulated jet flows expand as they emerge from the launch cylinder, then refocus. This repeats along the jet quasi-periodically within the head region, leading to a series of steady and stationary, successively weaker ``recollimation shocks'' (analogous to ``shock diamonds'' in engineering contexts) that extend down each jet (e.g.,~\cite{norman82,perucho13,bodo18}) (see also, e.g.,~figure \ref{earlyvslice}, where these features appear as jet constrictions), In the synchrotron images in figures \ref{t015_obs}, \ref{t045_obs}, \ref{t135_obs} and \ref{t135_pol} the recollimation features show as ``jet hotspots'' in the source head region. The recollimation shocks closest to the source can be moderately strong. In the reference simulation the innermost recollimation shock Mach numbers, $\mathcal{M} > 3$, with $\mathcal{M} \sim 6$ on the jet axis. DSA in the strongest shock sections harden the spectra of CRe populations passing through them (see \S \ref{partdyn}){\footnote{Recall that the distribution slope of the CRe population emerging from the launch cylinder corresponds to DSA from a $\mathcal{M} = 3$ shock.}}. Shocks further down the jets in this case, where the jet flow begins to deflect, are too weak (mostly $\mathcal{M} < 1.5$) to lead to significant DSA. We comment in passing that, although recollimation shocks are a common, natural occurrence in jet flows, they are typically associated with transitions within ambient conditions. In this simulation that transition comes at the end of the launch cylinder, while the analogous transition in a real RG would be on much smaller scales. Such a transition in the scales simulated here could in principle relate, for example, to jet emergence from the host galaxy ISM. We also mention that superficially similar features on the scales relevant here might result from time variations in the jet flow velocity, although those would not be stationary.

\subsubsection{NAT Phase Tail Dynamics}
\label{sec:tails}
Inherent to the nature of NAT tail formation, much of the AGN plasma becomes mixed with ambient, ICM plasma. That is evident in our reference simulation from jet color images in figures \ref{earlycolor} and \ref{wagfig}. Mixing comes about through a combination of three processes, beginning with intermittent jet disruption events that rapidly deposit the content of pinched jet segments. In addition to direct jet-ICM mixing from that, the turbulence driven by these events mixes plasma that continues to exit the termini of the active jets.  Consequently, most mixing takes place in what we called the disruption zone, and most of the jet material is in volumes with intermediate jet color values, $C_j \sim 0.3 - 0.7$. Between disruption events, jets burrow further into the tails (e.g.~figure \ref{wagfig}), injecting momentum, magnetic flux and CRe. In the process they also populate the tails with clumps of more pristine jet plasma ($C_j \sim 1$). In addition, even as they propagate through quiescent environments, the jets entrain some ambient plasma through their boundaries. In our simulations that last entrainment is largely a consequence of numerical diffusion, although analogous, real jets are thought to entrain ambient plasma by, for example, jet surface instabilities \citep[e.g.,][]{bicknell94,laing08}. 

Together these processes cause the tails to become highly heterogeneous blends of ICM and jet plasma. In our reference simulation the ICM is unmagnetized and free of CRe, so these mixing processes are essential to seeding the NAT tails with magnetic fields and CRe responsible for its radio synchrotron emissions.
\begin{figure}[ht]
    \centering
    \includegraphics[width=0.9\textwidth]{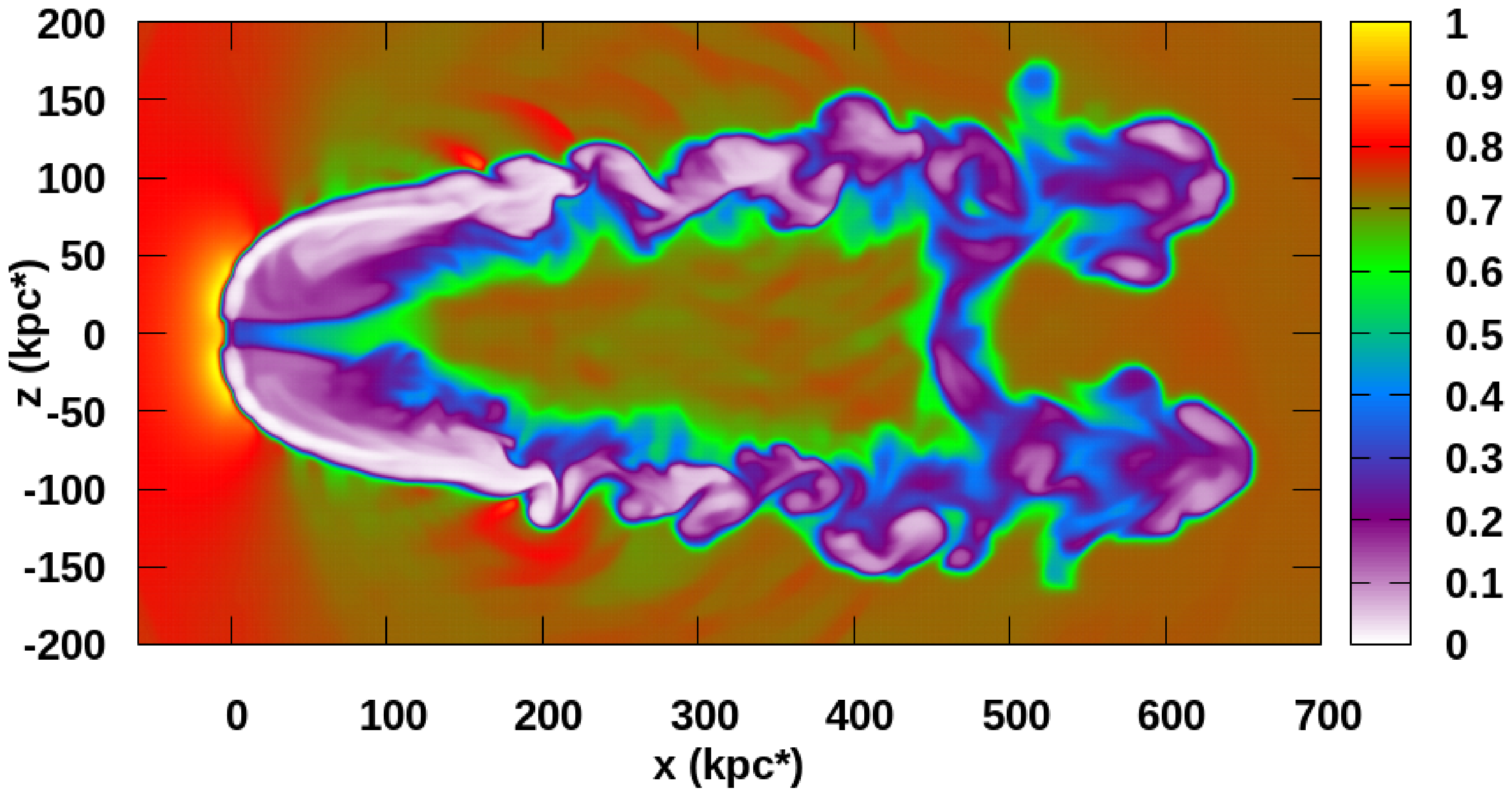}
    \caption{Projected mass density at $t = 541.1$~Myr$^*$ through a slab 15~kpc thick centered in $y$ on the simulation midplane. The projected mass density is scaled to the maximum value. In those units the undisturbed background ICM projected mass density $\approx 0.71$. The $^*$ in labels serves as a reminder that the nominal length (and time) scales can be increased or decreased by adjustment of the bending length, $\ell_b$.}
    \label{TailsRho}
\end{figure}
The mass distribution in the NAT tails is also highly heterogeneous. This is evident in figure \ref{TailsRho}, which displays the projected mass density along $y$ through a 15~kpc thick slab centered on $y = 0$, just before the AGN jets turn off. That heterogeneous density distribution will play a prominent role when, in a subsequent paper in preparation \cite{on19b}, we study simulated interactions between this NAT and cluster merger shocks. 

The lowest projected density regions visible in figure \ref{TailsRho} highlight tail volumes of mostly jet material ($C_j \sim 1$), and reveal that, well downwind from where the jets become unsteady, substantial tail volumes are dominated by identifiable pinched off jet segments that here project areas $\gtrsim10$~kpc wide and $\sim50$ - 100~kpc long. As one would expect, these features show obvious distortions derived from both jet flapping episodes and tail turbulence. On the other hand there is a generally positive column density gradient along each tail that reveals ongoing mixing between jet and ICM plasma. Also in figure \ref{TailsRho} we can see a bridging channel between the two tails that is the outcome of the initial, transient plume described in \S \ref{sec:early}. In addition and downstream (to the right) of that feature are the remnants of the early jet disruption underway in figures \ref{earlycolor}, \ref{earlyvslice} and \ref{wagfig}. By the time of figure \ref{TailsRho}, iC radiative cooling has caused these early disruption features to become virtually invisible through radio emissions.
\begin{figure}[ht]
    \centering
    \includegraphics[width=0.6\textwidth]{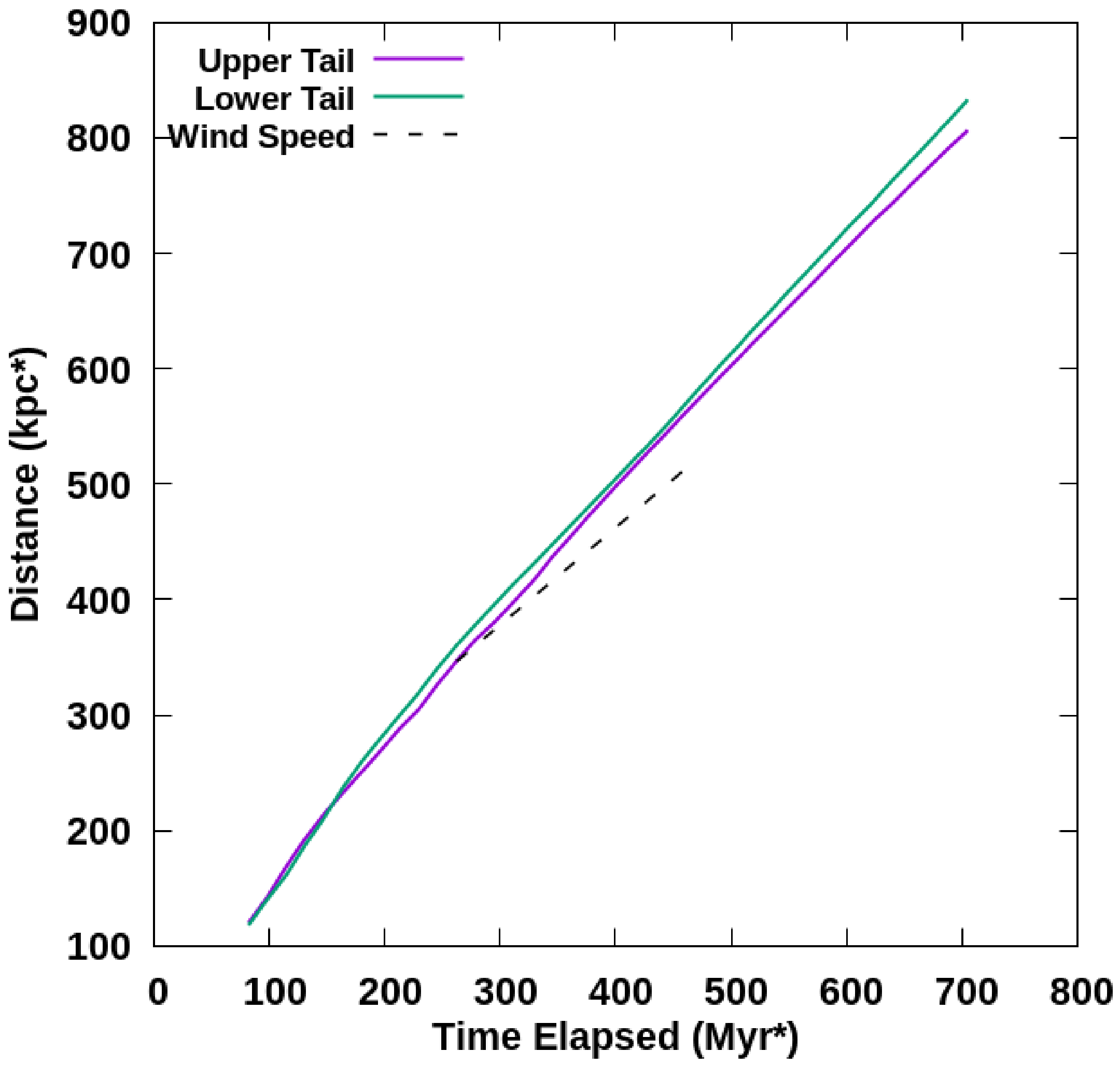}
    \caption{Distance (along $x$) vs time from the AGN to the end of each tail, compared with $x = v_w ~t$. The~$^*$ in labels serves as a reminder that the nominal length (and time) scales can be increased or decreased by adjustment of the bending length, $\ell_b$.}    
    \label{TailExpfig}
\end{figure}

The density distribution evident in Figure \ref{TailsRho} also provides a simple way to understand the point made in \S \ref{sec:early} that the NAT tails extend downwind at speeds greater than that of the driving wind, $v_w$; the tail plasma is not passively advected in the wind. The underlying physics alluded to there was that in the head regions of the source jets extract momentum from the wind as they are deflected downwind, then return that momentum downwind into the tails. The net momentum transfer should balance on average. But, since the density of the tail plasma is less than the density in the wind itself, the resultant downwind velocity of the tail plasma exceeds that of the wind. Quantitatively, the difference should depend on both the density contrast and the effective cross section ratio, so is not trivial to calculate.  The outcome in our reference simulation is illustrated in figure \ref{TailExpfig}, where we can see that the (mean) tail extension rate is $\sim 10$\% faster than the background, wind flow{\footnote{For analyses of this kind we define ``tail plasma'' as plasma with jet color $0.01<C_j < 1$}}. As already mentioned, for a NAT formed in a post shock wind \citep[][]{nolt19b}, significant consequences of this enhanced tail extension rate would likely include long term interactions between the initiating shock and the NAT tails, possibly leading to transfer of CRe from the tails to the shock. More generally, this would enable the NAT tails to overtake extraneous ICM features, such as shear layers akin to ``sloshing fronts'' and, conceivably, ``light them up'' by transferring CRe. Such events might account for some sharp bends that have been noted in NAT tails \citep[e.g.,][]{hoang17}.

To understand tail morphologies as illustrated in figure \ref{TailsRho}, it may be helpful to point out that, once the tails evolve into the NAT phase, they are roughly in total pressure balance with their surroundings. As such, the long term widths of the tails are largely determined by turbulent eddy motions, with some bias due to residual lateral divergence of the jet trajectories (see the Appendix).

We mention for use below that analysis of  2nd order velocity structure functions in the tails indicates that tail turbulence has a characteristic velocity of about half the wind speed ($\delta v \sim v_t \approx 0.5v_w = 450\kms$) with an outer scale roughly the jet bending length, $\ell_b \approx 35$ kpc. The eddy turnover time associated with this scale in this simulation is, therefore, $\tau_{eddy} \sim \ell_b/v_t  \sim 80$ Myr . We do emphasize, however, that turbulence in the tails is neither homogeneous, nor steady.

Finally, we briefly comment on subsequent evolution of the simulated NAT following termination of AGN activity. \cite{on19b} present extensions of the evolution of the NAT discussed here, in which it is overrun, one tail after the other by an ICM shock, and where, during the encounter between the shock and the first tail, jet activity is terminated. We refer readers to \cite{on19b} for details. Depending on the Mach number of the shock in those encounters several 10s of Myr pass after AGN activity ceases and the second tail has shock impact. That is long enough to reveal the general character of post activity dynamics, as well as to allow significant radiative cooling of the relevant CRe population. The evolution of the second tail (and the jet that created it) until shock impact on the second tail is virtually the same as that for both tails in the absence of the shock. Very briefly, immediately after the jet launch activity ceases, recently launched jet plasma clears the jet channel into the tail. The time required is very roughly $\sim \rm{a~few} \times \ell_b/v_j$, which obviously depends on both the jet velocity and the length, $\ell_b$. For the parameters of the present simulation, the time required is only a few Myr. From that time no additional energy, magnetic flux or CRe would be injected into the tails from the AGN, so the tails would appear in isolation and their flows, including turbulence, would decay. Accordingly, their synchrotron emissions would fade, as well, typically on timescales $\la 100$ Myr. Those times are short enough that the likelihood of observing radio bright tails that are detached but not somehow ``reactivated'' (perhaps by a shock) is relatively small.

\subsection{Magnetic Field Dynamics} \label{MagDyn}
The magnetic field in our jet launch cylinder is toroidal and increases linearly in strength towards the cylinder perimeter with constant peak value, $B_0 \approx 1 \mu$G. Upon emergence from the launch cylinder the magnetic field is governed by the magnetic induction equation. In the ideal MHD limit, that can be written

\begin{equation}
\frac{\partial \vec{B}}{\partial t} = \vec{\nabla} \times \left( \vec{v} \times \vec{B} \right).
\end{equation}

Applying standard vector identities, $\nabla\cdot\vec{B} = 0$, and the mass continuity equation, $\nabla\cdot\vec{v} = -D\ln{\rho}/Dt$, this can be expressed simply in terms of Lagrangian time derivatives, $D/Dt = \partial/\partial t + \vec{v}\cdot\nabla$, as

\begin{equation}
    \frac{D\vec{B}}{Dt} = \frac{D\ln{\rho}}{Dt} \vec{B} + (\vec{B}\cdot\nabla)\vec{v}.
    \label{eq:induct}
    \end{equation}

\begin{figure}[ht]
    \centering
   \includegraphics[width=\textwidth]{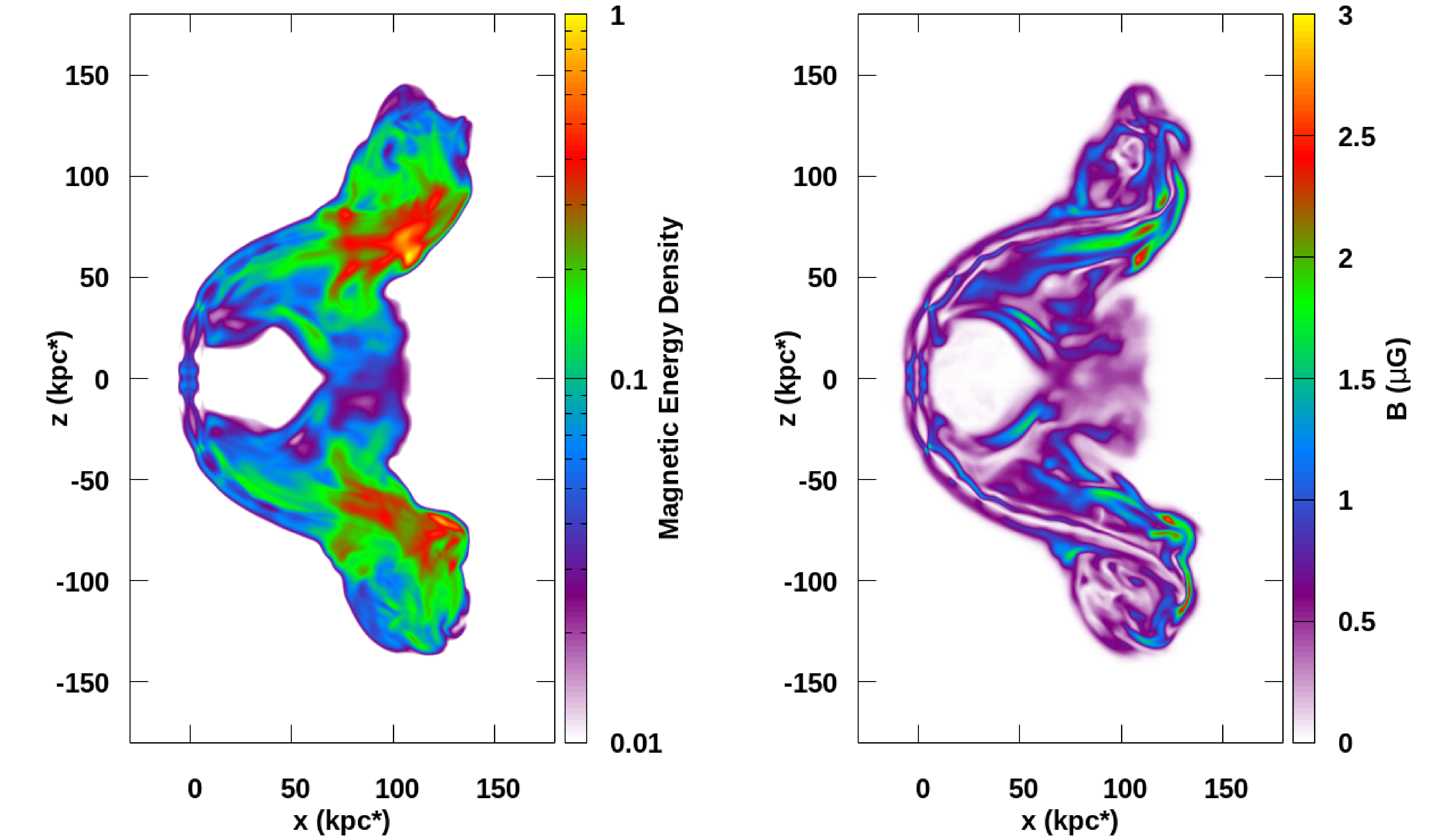}
\caption{Magnetism at $t = 98.4$~Myr. Left: Projection along y of magnetic energy density (arbitrary units). Right: Plane ($y = 0$) slice of magnetic field strength ($\mu$G). The associated distribution of jet color, $C_j$, is displayed on the right side of figure \ref{earlycolor}. The $^*$ in labels serves as a reminder that the nominal length (and time) scales can be increased or decreased by adjustment of the bending length, $\ell_b$.}
\label{t30_B}
\end{figure}

The first RHS term in equation \ref{eq:induct} accounts for plasma compression, while the second accounts for stretching of magnetic flux lines. Although both terms contribute to the evolution of the magnetic field in our simulation, the second, stretching term is usually predominant, especially during jet disruption. The distribution of magnetic energy in figure \ref{t30_B} at $t \sim 100$~Myr demonstrates the importance of this dynamic. The fast moving jets are put under great dynamical stress when their velocities change abruptly, and these stresses convert some of the jet's kinetic energy into magnetic energy. During the early dynamical phase, the rapidly extending jets experience larger stresses than later on, so the plumes that cap the ends of the tails contain strongly magnetized plasma (see right side of figure \ref{earlycolor}, figure \ref{t30_B}).

In the ideal MHD limit the force equation of relevance to us, including the Maxwell stress term in rationalized units, is

\begin{equation}
    \rho \frac{D\vec{v}}{Dt} = \rho\left(\frac{\partial \vec{v}}{\partial t} + \vec{v}\cdot\nabla \vec{v}\right) = -\nabla P_g + \vec{J}\times\vec{B} = - \nabla( P_g + P_B) + \left(\vec{B}\cdot\nabla\right)\vec{B}.
\label{eq:Euler}
\end{equation}

The last RHS term in equation \ref{eq:Euler} accounts for magnetic tension. The relative importance of gas pressure and magnetic pressure is generally expressed by $\beta_p = P_g/P_B$, while the relative role of magnetic tension compared to plasma inertial stresses, $\propto \rho v^2$, is expressed through the Alfvenic Mach number, $M_A = v/v_A = \sqrt{5\beta_p/6}~ M$, where $v_A = B/\sqrt{\rho}$ and $M = v/c_s$ is the sonic Mach number. Only when both $\beta_p$ and $M_A$ are large can we be confident that Maxwell stresses are subdominant.

\begin{figure}[!ht]
    \centering
    \includegraphics[width=0.75\textwidth]{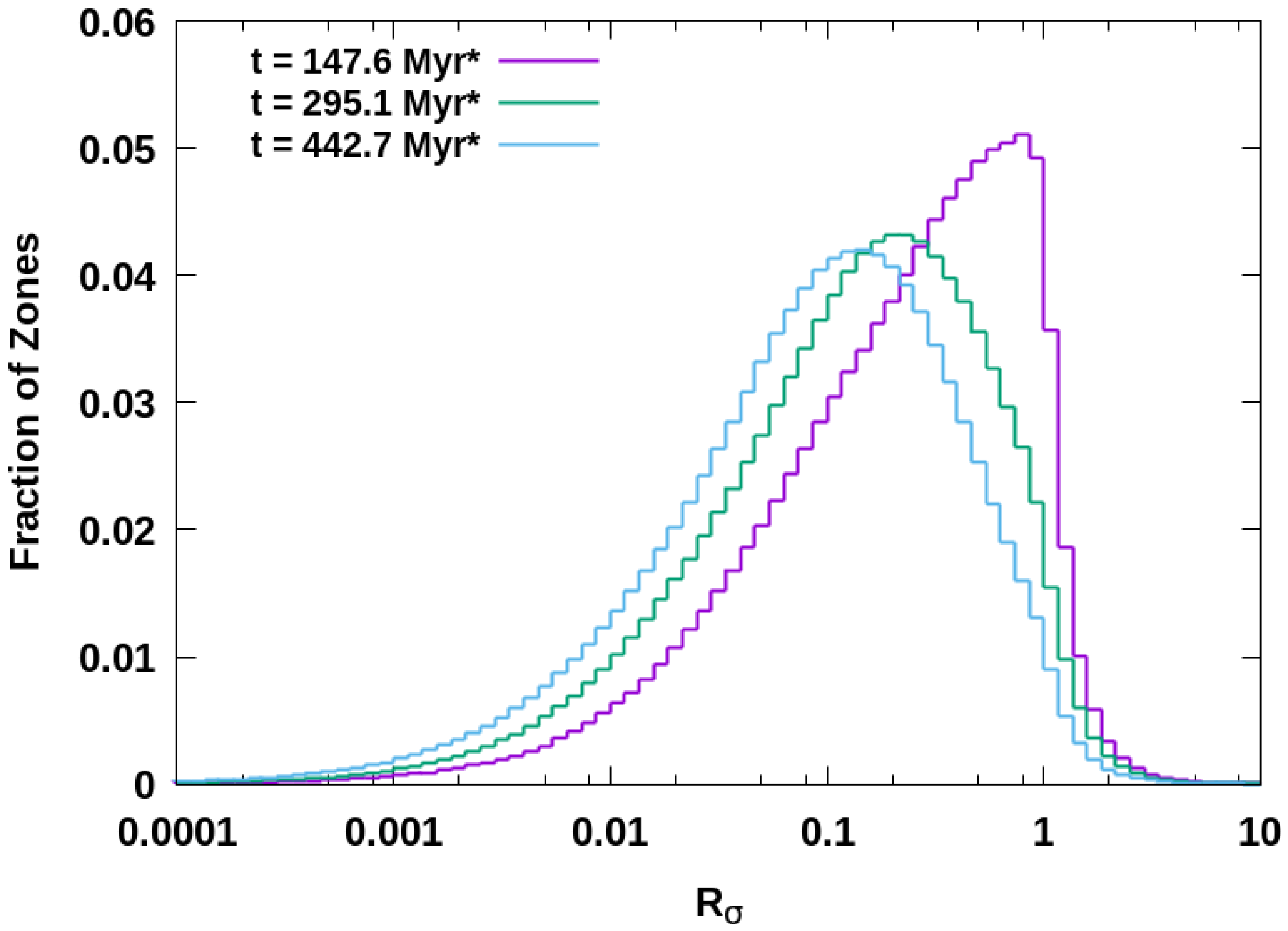}
    \caption{Log-linear histogram of the distribution of $R_\sigma$ in computational cells with significant jet mass fraction, $C_j \geq 1$\%, at three times during evolution of the reference NAT.}
    \label{RstressHist}
\end{figure}

The jet plasma is initialized with a moderately large fiducial plasma beta, $\beta_{pj} = 25$. The analogous Alfvenic Mach number of the emergent jet is $M_{Aj} \sim 14$. Consequently, magnetic fields in the emergent jets are minor dynamical contributors. Nevertheless, as the jets propagate and disperse, regions quickly develop that are obviously influenced by Maxwell stresses nearly as much or more so than hydrodynamic stresses, particularly near regions of jet disruption. To explore this more fully we compute in each jet or tail simulation cell, $R_{\sigma}$, the ratio of Maxwell to hydrodynamical stresses; i.e.,

\begin{equation}
R_\sigma = \frac{\left| \vec{J} \times \vec{B} \right| }{\left| \nabla P_g + \rho\left(\vec{v} \cdot \nabla\right)\vec{v} \right| }
= \frac{\left|\left(\vec{B}\cdot\nabla\right)\vec{B} -\nabla P_B \right| }{\left| \nabla P_g + \rho\left(\vec{v} \cdot \nabla\right)\vec{v} \right| }.
\end{equation}

Values of $R_{\sigma}\ga 1$ indicate strong Maxwell stress influence. As plasma emerges from the jet launch cylinder,  $R_{\sigma} \la 1/M_{Aj}^2 \la 0.01$. Figure \ref{RstressHist} shows the full distribution of $R_\sigma$ at three times in cells that have $C_j \geq 0.01$ (that is, at least 1\% of the mass in the cell originated in one of the jets, which are the only sources of magnetic field in the simulation). The total number of cells included in the histograms increases from roughly 1\% to 4\% of the simulation domain over the time span sampled. The $R_{\sigma}$ distribution at the earliest time shown, $t = 147.6$~Myr, when the NAT phase is becoming clearly established (see figure \ref{wagfig}), resembles a log-normal curve for low values of $R_\sigma$ and peaks around $R_\sigma \approx 0.8$. The fraction of cells with $R_\sigma \geq 0.3$ in the distribution at this time is $\approx 45\%$ of those included, although there are relatively few with $R_{\sigma}\geq 1$. Throughout the early NAT phase, we find a similar distribution in $R_{\sigma}$ that peaks just below one, then drops sharply at higher $R_{\sigma}$ values. Such a distribution is consistent with a plasma that is threaded with magnetic fields that are stretched and strengthened by fluid motions (cf. equation \ref{eq:induct}) until Maxwell stresses are strong enough that they inhibit those motions and stifle further field amplification (cf. equation \ref{eq:Euler}); i.e.,~the magnetic field amplification saturates. Closer examination of the stresses shows that magnetic tension generally plays a larger dynamical role than magnetic pressure. At later times, as the NAT structure becomes more extended, but also as jet disruptions influence relatively less of the tail structures, the Maxwell stresses play less significant roles overall. This is evident from the $R_{\sigma}$ distributions at $t = 295.1$~Myr and at $t = 442.7$~Myr shown in figure \ref{RstressHist}.

\subsection{Cosmic Ray Electron Dynamics} \label{partdyn}

As outlined in \S \ref{SimDet}, our simulation tracks a CRe population within the particle Lorentz factor range, $10 \la \Gamma_e\la 2\times 10^5$. We explore below synthetic synchrotron observations at frequencies $\nu \la 1$~GHz in the source frame (so, at z = 0.2, 20\% above the observed frequency). Those emissions in our reference NAT come primarily from regions with magnetic field strengths very roughly, $B \sim 1-10 \mu$G. Using standard synchrotron relationships ($\nu_{GHz} \sim \nu_c\sim 4\times 10^{-9} B_{\mu\text{G}} \Gamma_e^2$; ~$\nu_c \sim (3/2) \Gamma_e^2 \nu_{B}$, with $\nu_c$ the so-called critical synchrotron frequency and $\nu_B$ the nonrelativistic electron cyclotron frequency) brighter emissions in the band we model relate very roughly to $\Gamma_e \la 10^4$ \citep[e.g.,][]{rybicki79}. 

After they emerge from the launch cylinder our CRe can gain and lose energy in several ways. They can be accelerated via DSA during any sufficiently strong shock passage, based on behaviors laid out in \S \ref{SimDet}.  Briefly reiterating, within energy bins spanning about a factor 3, the CRe particle momentum distribution emergent from a shock is assigned a power law slope that is the flatter of the incident slope and that defined by test-particle DSA at that shock ($q = 4M_s^2/(M_s^2 - 1)$). However, in the reference simulation under discussion, the only shocks strong enough to influence the CRe are shocks within the jets near their launch cylinder. More broadly, the CRe are subjected to energy losses and gains from adiabatic expansion and contraction as well as losses by radiative cooling (synchrotron and iC off CMB photons). Since magnetic fields in our simulation are predominantly less than the radiatively equivalent field strength of the CMB ($B_\text{CMB} \approx 4.7\mu$G at z = 0.2), iC losses mostly, but not universally, dominate synchrotron losses in our simulation. The iC cooling time at redshift, z = 0.2, is \citep[][]{szin99},
\begin{equation} \label{eqn:CRage}
\tau_{rad} (\Gamma_e) \approx \frac{10^6}{\Gamma_e} ~ \text{ Myr}  = \frac{\tau_o}{\Gamma_e}.
\end{equation}
This translates for $\Gamma_e \sim 10^4$ very roughly as $\tau_{rad} \sim 100$~Myr, so of the same order as the dynamical evolution timescale of our NAT ($\sim\rm{a~ few} \times \ell_b/v_w \sim 100$~Myr) and to our estimated turbulent eddy time, $\tau_{eddy} \sim 80$~Myr. The jet speed, $v_j \approx 25$~kpc/Myr, is more than an order of magnitude faster than the wind speed, $v_w$, in our simulation, so, while AGN plasma remains inside coherent jet flows, it can traverse the NAT structures in only a few Myr. Even accounting for jet disruptions that limit this direct transport, fast jet transport accounts for the obvious, ongoing CRe refreshment described in \S \ref{sec:obs}. On the other hand, the complex CRe histories within the object also explain quite naturally why the local CRe electron distributions, $f(p)$, are quite heterogeneous, and mostly not well described as power laws over broad energy ranges.

We did not model 2$^\text{nd}$ order CRe reacceleration by NAT-tail turbulence (``Fermi II'' reacceleration) in our simulation. As noted above, tail turbulence is highly inhomogeneous and unsteady. Given that the efficiency of Fermi II reacceleration depends sensitively on details of how turbulent kinetic energy cascades to wave energies on scales orders of magnitude smaller than we consider in our simulations \citep[e.g.,][]{brunetti11,lynn14}, and that the timescales for the energy cascades are comparable to those associated with jet disruption events, meaningful Fermi II models in this context  would require considerable carer.

Sophisticated Fermi II modelling is beyond our present scope. We will address the issue in future work. For now we refer only to a simple, crude relationship between the large eddy timescale, $\tau_{eddy}\sim \ell_t/\delta v_t$ and the Fermi II reacceleration time based on so-called ``TTD'' resonance between the CRe and MHD fast-modes; namely, $\tau_{acc} \sim \tau_{eddy} (c/v_t) (1/\Phi)$. Again, $v_t$ is the outer scale turbulent velocity, while $\Phi>1$ is a model-dependent function that relates to the fast mode turbulent wave spectrum and the smallest dissipation scale of fast mode turbulence in the medium of interest \citep[e.g.,][]{bj14}. From this we expect that $\tau_{acc} >  \tau_{eddy}$, which reinforces our concern about the unsteady character of NAT-tail turbulence in evaluating the role of Fermi II acceleration in the current context.

\section{Synthetic Radio Observations}
\label{sec:obs}

We computed and analyzed synthetic radio synchrotron emissions of the reference NAT source in order to link the dynamical behaviors outlined above with associated observable properties. Spectral intensity, $I_{\nu}$, and associated polarization and spectral index, $\alpha_{\nu 1,\nu 2} = - log(I_{\nu 1}/I_{\nu 2})/log(\nu 1/\nu 2)$, maps were obtained by integrating spectral emissivity along lines of sight through the simulated volume. Emissivities, including linear polarization, were computed at a given frequency, $\nu$, utilizing full 3D vector information about the magnetic field and integrating the synchrotron emission kernels, $F(\nu/\nu_c)$ and $G(\nu/\nu_c)$, over $\nu_c$, where $\nu_c = (3/2)\Gamma_e^2 \nu_{B\perp}$, and $\nu_{B\perp} = \nu_B \sin{\theta}$, with $\theta$ the angle between the local magnetic field and the line of sight \citep[e.g.,][]{rybicki79}. If the CRe distribution (assumed isotropic) is a pure power law, $f(p) \propto p^{-q}$, over a broad range [$p_1,p_2$], and $\nu_{c1} \ll \nu \ll \nu_{c2}$, where $\nu$ is the frequency of observation while $\nu_{c1}$ and $\nu_{c2}$ are the critical synchrotron frequencies for $p_1$ and $p_2$, then $\alpha = (q - 3)/2$. Under similar conditions, accounting for shifts in $\nu_c$ at fixed $\nu$, $I_{\nu} \propto \nu_{B\perp}^{(q-1)/2}$. As CRe in our simulation first emerge from the jet launch cylinder with $f(p) \propto p^{-4.5}$ over a wide momentum range. So, at radio frequencies we model here, the synchrotron emission has an index, $\alpha = \alpha_0 = 0.75$, while the nominal intensity-magnetic field relation is $I_{\nu} \propto \nu_{B\perp}^{1.75}$. However, due to a combination of non-adiabatic gains and losses (predominantly radiative losses), the CRe distributions in our simulated NAT are not generally well-described as power laws over relevant energy ranges, so it is important to compute the synchrotron spectrum more carefully, as described above. Of course, as is well known, when CRe populations cool radiatively, $f(p)$ steepens from the top down so the associated synchrotron spectra steepen most rapidly at higher frequencies; i.e.,~the spectra ``age.'' 

Radiative cooling as the CRe propagate down the tails, obviously plays a major part in determining the visible lengths of the tails as radio sources. In reality, the tail lengths, and their lengths in comparison to their widths determined in this way, will depend on such details as the actual physical magnetic field intensity and on the actual redshift of an observed NAT. In addition, if Fermi II reacceleration  due to turbulence in the tails is, indeed, fast enough to play a role, this would obviously tend to increase observed lengths compared to what we find.

Synthetic synchrotron images, as well as associated spectral and polarization maps of our reference NAT at reference frequencies, 150~MHz, 325~MHz, 600~MHz, 950~MHz and 1.4~GHz in the observer frame{\footnote{These would become 180~MHz, 390~MHz, 720~MHz, 1.14~GHz and 1.68~GHz in the source frame for z = 0.2.}} are shown in figures \ref{t015_obs}, \ref{t045_obs}, \ref{t135_obs}, and \ref{t135_pol}. To facilitate connection of emission features to dynamical features, the simulation $x$-$z$ plane was placed in the plane of the sky.  Bilinear-averaged observing beams $0.895$~kpc diameter (spanning 1.79 grid cells) were applied. From these images we also computed integrated radio fluxes and spectra reported in figures \ref{flxevo}, \ref{slopevo} and \ref{IndivSpectra}. The polarization calculations did not include Faraday rotation. They would not be meaningful for this simulation given our simplifying assumption that the ICM (including ICM mixed into the NAT) was unmagnetized.

\subsection{Analysis of Synthetic Images}
\label{sec:synchimages}

We examine here synthetic intensity and spectral index ($\alpha$) maps from three snapshots in time spaced through this RG's evolution; two from the early phase when the NAT structure is still being defined (\S \ref{sec:early}) and the other from well into the NAT phase (\S \ref{sec:tails}).

\begin{figure}[ht]
    \centering
    \includegraphics[width=\textwidth]{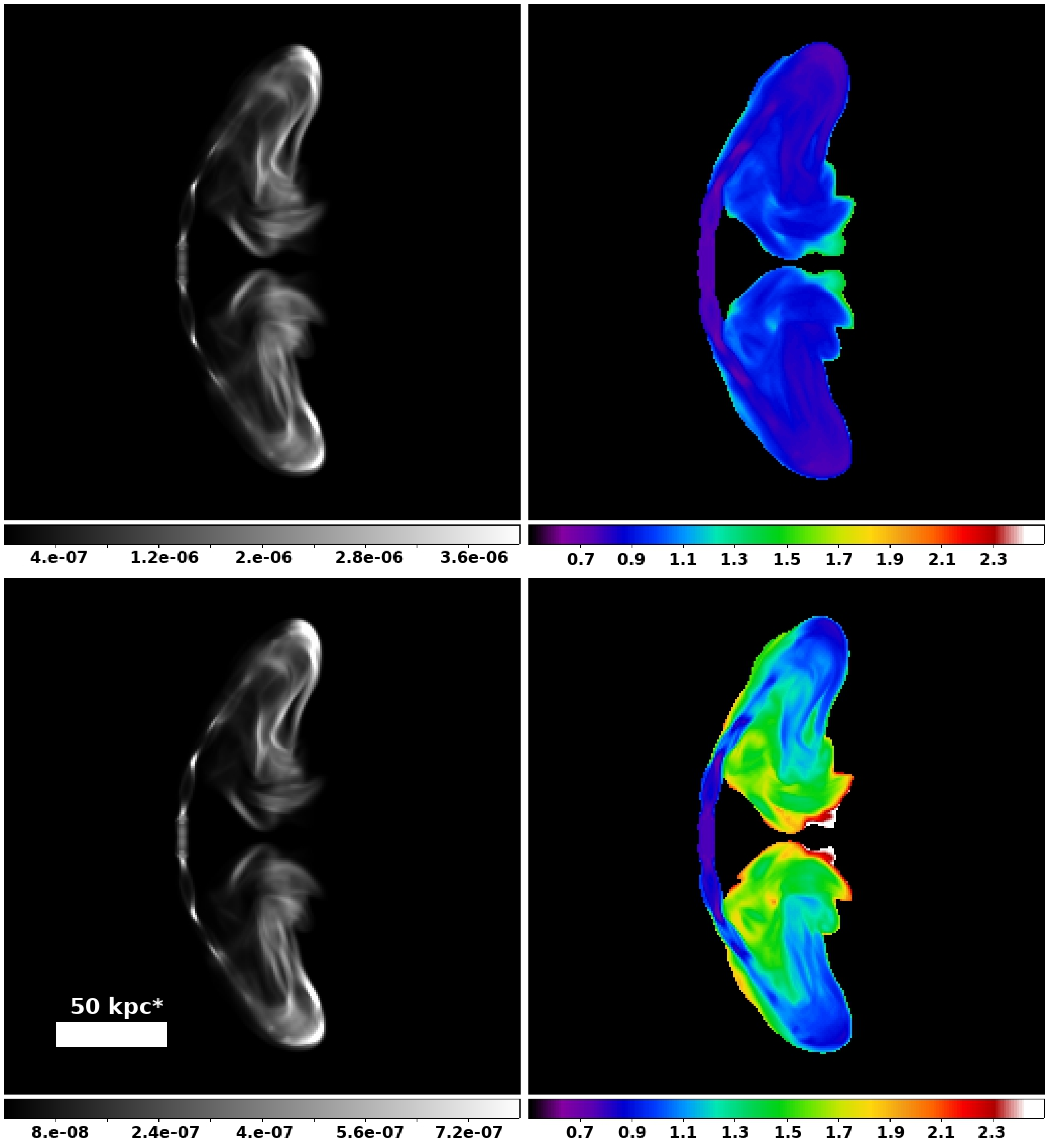}
\caption{Radio images at $t = 49.2$ Myr$^*$. Top Left: $I_{\nu,150}$. Bottom left: $I_{\nu,950}$. Top right: $\alpha_{150,325}$. Bottom right: $\alpha_{950,1400}$. Intensities are in arbitrary units. Brightest regions have been slightly saturated to reveal fainter features. Multiple jet hot spots in the head on either side of the launch cylinder correspond to recollimation features mentioned in the text. The $\alpha$ maps include only pixels having $I_{\nu}$ within a factor 500 of the brightest pixels at the higher frequency. The $^*$ in labels serves as a reminder that the nominal length (and time) scales can be increased or decreased by adjustment of the bending length, $\ell_b$.}
\label{t015_obs}
\end{figure}
The 150~MHz and 950~MHz intensity images in figure \ref{t015_obs} from $t = 49.2$~Myr show how our RG appears from this observer's perspective relatively early in its evolution (see figure \ref{earlyvslice} for associated dynamical structures). Depending on spatial resolution (and orientation relative to the observer), the apparent radio jets and plumes could be interpreted as the gently bent lobes of a WAT RG. On the other hand the high surface brightness at the ends of the jets (``hot spots'') is characteristic of a morphological class II Fanaroff-Riley radio galaxy. Viewed with the plane containing the tails significantly out of the plane of the sky it is likely this object would be so-labeled.  So, at this stage the future NAT appearance of this RG is not at all obvious. The first two pairs of jet recollimation shocks mentioned in \S  \ref{sec:natjet} are evident as bright spots in the intensity images of figure \ref{t015_obs}. In addition,  flattening of both $\alpha_{150,325}$ and $\alpha_{950,1400}$ at the first (stronger) recollimation shock pair, and of $\alpha_{950,1400}$ at the second such shock pair provide evidence for DSA in those shocks ($\alpha_{950,1400}$ is more sensitive in this context, since it communicates properties of higher energy CRe.) It is notable that these shock features are the only locations in the images at this time that rival the brightest structures at the ends of the jets. Those latter features correspond to highly bent, ``soon-to-be-disrupted'' jet segments discussed in \S \ref{sec:early}. In addition, the magnetic fields in those regions have been strongly enhanced by stretching ($B> 3\mu$G).

Because jet propagation is fast ($v_j \approx 25$~kpc/Myr), the CRe populations in the brightest regions of the $t = 49.2$~Myr snapshot are quite young. Radiative cooling influences are negligible at the lower observing frequencies, especially because the magnetic fields in those regions are relatively large, meaning that at fixed observing frequency, the CRe that dominate the emission have lower $\Gamma_e$. Even observed near 1~GHz iC radiative spectral aging of the synchrotron spectrum at this time is modest where B is stronger and due mostly to depletion of higher energy CRe with $\nu_c$ above the observed frequencies. Accordingly, in these regions, the lower frequency spectral index, $\alpha_{150,325}$, is close to the $\alpha \approx \alpha_0 =0.75$ expected at jet launch. Even at the higher frequencies in these bright regions $\alpha_{950,1400} \la 0.9$ at this time. Lower surface brightness regions in the image of $I_{\nu,950}$ from figure \ref{t015_obs} with $\alpha_{950,1400} \gtrsim 1.1$ come from diffuse, less directly propagated portions of the plasma plumes stripped from the young jets as described in \S \ref{sec:early}. These regions also have weaker magnetic fields, and thus emissions correspond to larger $\Gamma_e$ with somewhat more ``aged'' CRe than in the brightest places. At 1.4~GHz, the contributing CRe in weaker field regions ($B < 1 \mu$G) have iC cooling timescales $\tau_{rad} \la 30$~Myr, so the high frequency spectral map reveals patterns of CRe energy evolution in such regions reflecting the history of jet plasma deposition. This accounts for the clear $\alpha_{950,1400}$ gradient within the bridge between the tails. Specifically, the spectrum steepens from the youngest CRe near the jets to older CRe at the perimeters of the plumes, now near the central plane. 

\begin{figure}[ht]
    \centering
    \includegraphics[width=\textwidth]{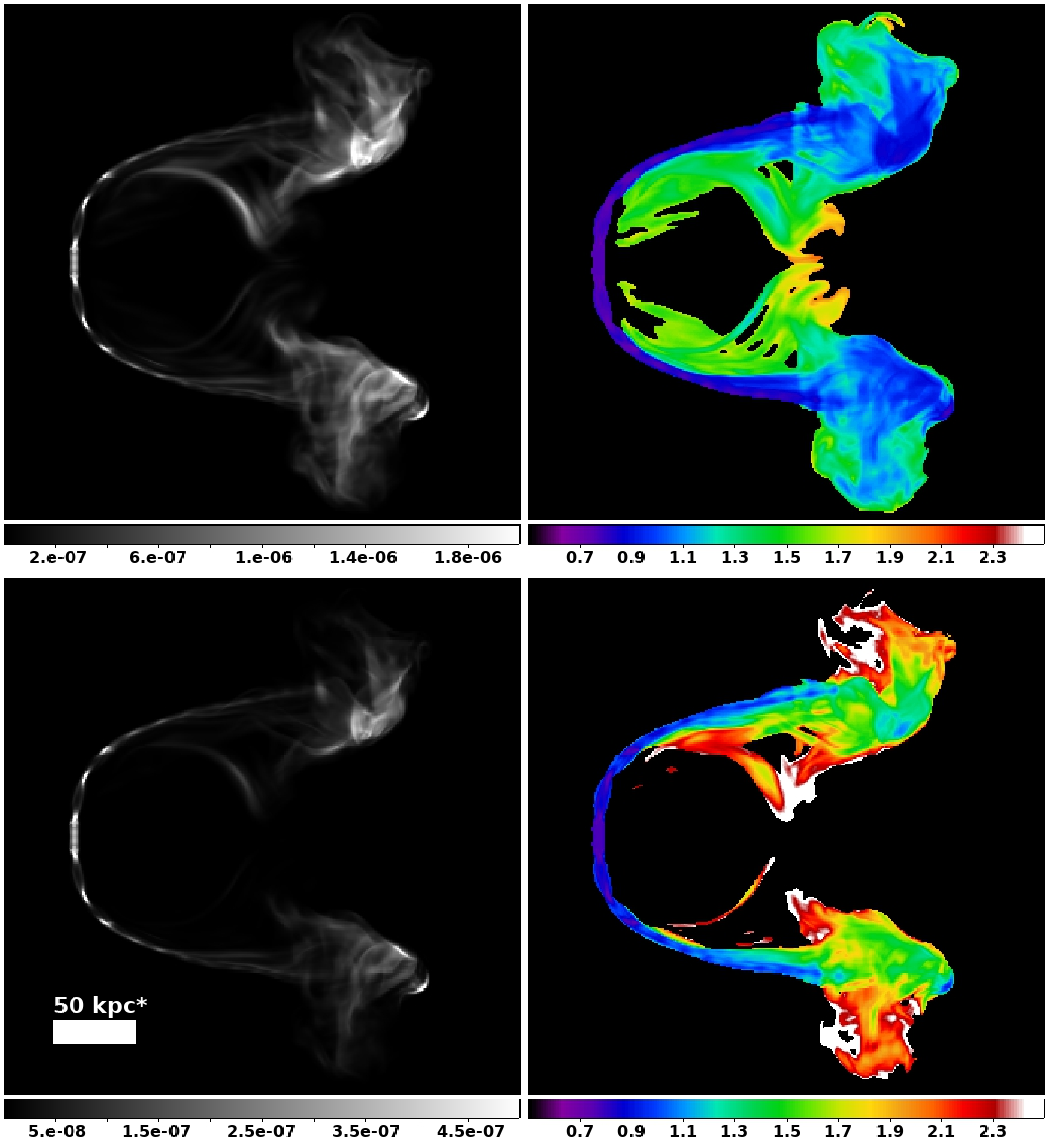}
\caption{Same as figure \ref{t015_obs} at $t = 147.6$ Myr$^*$. The $^*$ in labels serves as a reminder that the nominal length (and time) scales can be increased or decreased by adjustment of the bending length, $\ell_b$.}
\label{t045_obs}
\end{figure}

By the time of figure \ref{t045_obs} ($t = 147.6$~Myr), near the end of the transition out of early source evolution, the basic NAT structure that characterizes this object at later times is evident (figure \ref{wagfig} offers some insights to the dynamical state around this time). Even though it would likely be identified by observers as a NAT, the synchrotron source at $t \approx 150$~Myr has some observable properties that distinguish it from its appearance at significantly later times. For example, CRe in the bridge between the tails that bounds the head wake (e.g.,~figure \ref{t30_B}), still produce observable synchrotron emissions at 150~MHz. The brightness falls substantially later on. Even at this time the bridge feature has begun to fade at higher frequencies, since the CRe there have not been refreshed in over 100~Myr. There is a rim-like arc bounding the upwind side of this plasma bridge that is still visible, although $\alpha_{950,1400} > 1.5$, so it is fading seriously.

The brightness distributions in figure \ref{t045_obs} at both high and low frequencies are dominated by the downwind extremes of the source; i.e.,~the nascent tail extremes, where jet plasma has been relatively recently deposited by the jet disruption event that began around $t \sim 100$~Myr (see figures \ref{earlycolor}, \ref{wagfig}). Not only are the CRe distributions there relatively young, but, as pointed out above, the magnetic fields in those regions tend to be strongly amplified by stretching.  The brightest patches in each tail, which have relatively flatter $\alpha$, correspond to collections of strongly magnetized filaments that are obvious in figure \ref{t30_B}. Indeed, at high spatial resolution, such as in figure \ref{t045_obs}, the intensity distributions of the tails is very filamentary, reflecting the magnetic field topology. This behavior continues through the NAT phase of the source evolution.

\begin{figure}[ht]
    \centering
    \includegraphics[width=\textwidth]{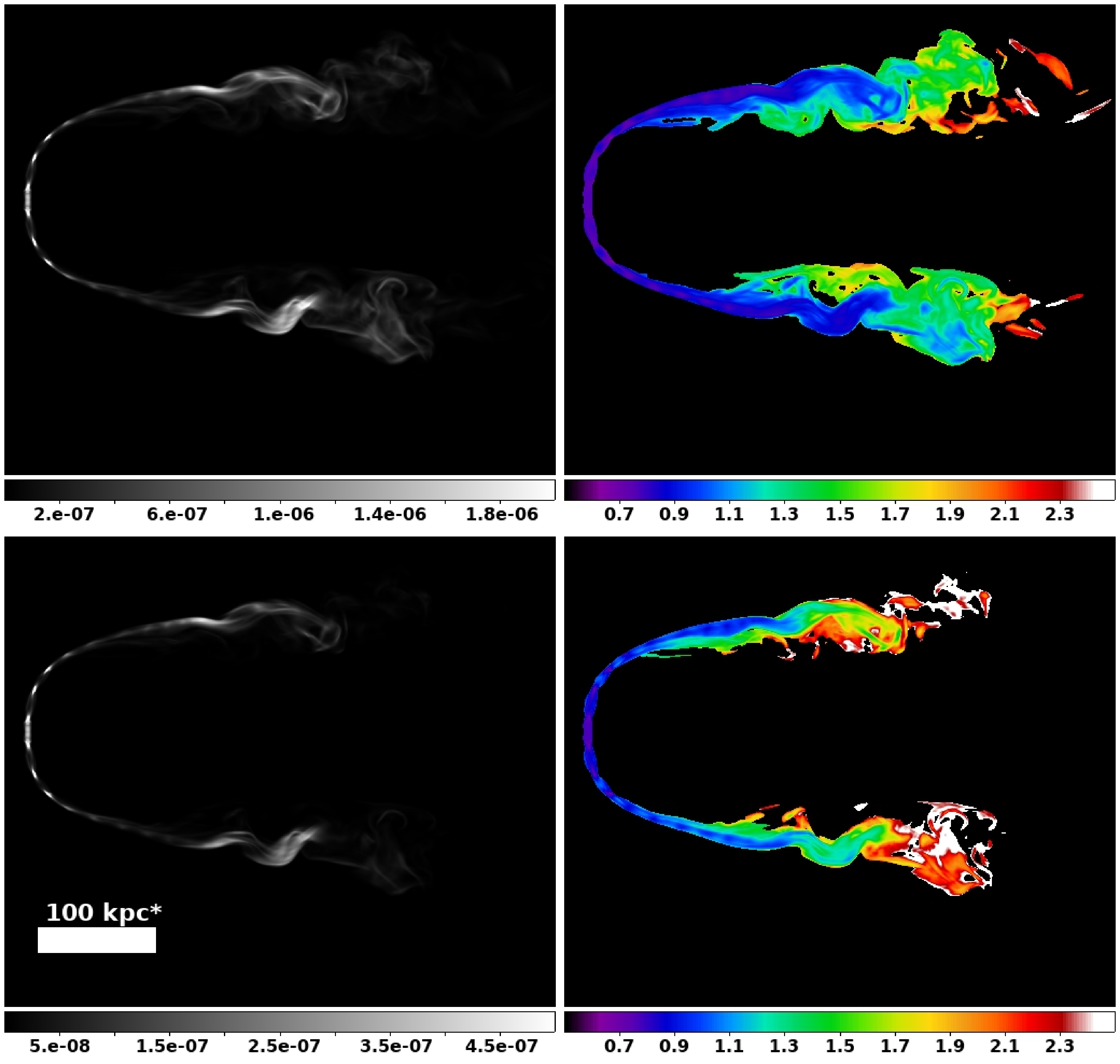}
\caption{Same as figure \ref{t015_obs} at $t = 442.7$~Myr$^*$. The $^*$ in labels serves as a reminder that the nominal length (and time) scales can be increased or decreased by adjustment of the bending length, $\ell_b$.}
\label{t135_obs}
\end{figure}

Figure \ref{t135_obs} shows radio images of the NAT 442.7~Myr into the simulation, well into the NAT phase of evolution, and while the jets remain active. The jets have very recently undergone a significant disruption event. The brightness distributions and spectra strongly reflect the recent history of local AGN plasma, including CRe and magnetic fields. The recollimation features in the jets in the head region, with their fresh CRe at relatively high densities and compressed magnetic fields (cf. figure \ref{t30_B}) are distinct.  On the other hand, the tail-bridging ``plumes'' as well as more extreme tail sections that were apparent in figure \ref{t045_obs} have not had their CRe replenished for a very long time, so have essentially faded away from visibility as a result of iC cooling, adiabatic expansion and magnetic field  relaxation. Even what we called in  \S \ref{sec:early} the ``disruption zone'' just downwind of the head that was once prominent, is relatively faint at this time. Disruptive jet actions that help to rejuvenate emissions have moved somewhat downwind. Indeed, the brightest radio emissions at this time are mid-tail and correspond to plumes of plasma from the most recent jet disruption events. The relatively flat spectra in those regions visible in images of $\alpha_{150,325}$ and $\alpha_{950,1400}$ from figures \ref{t135_obs} attest to the youth of those CRe.

\begin{figure}[ht]
    \centering
    \includegraphics[width=\textwidth]{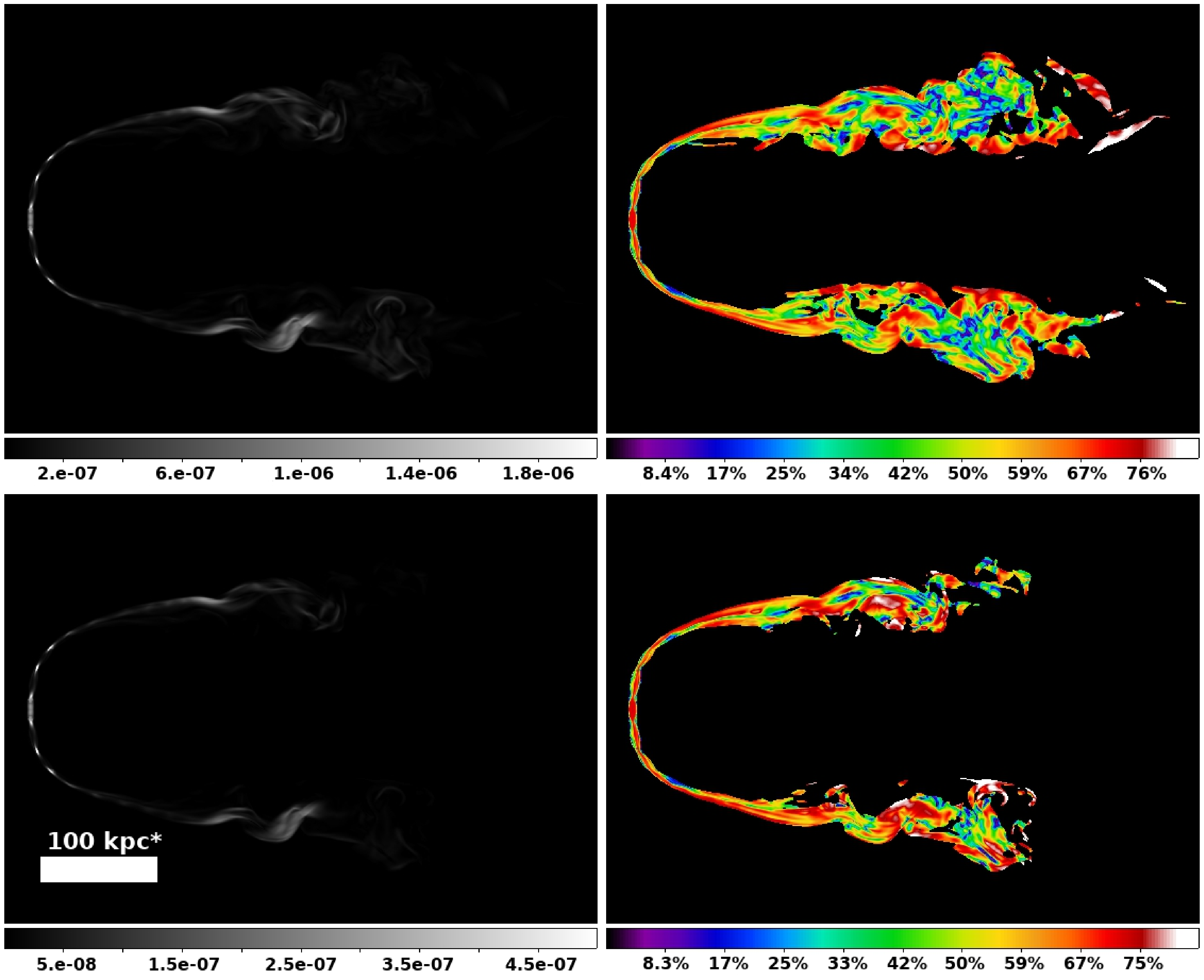}
\caption{Polarized radio emission at $t = 442.7$~Myr$^*$. Top left: polarized intensity at $\nu$ = 150~MHz. Top right: fractional polarization at $\nu$ = 150~MHz. Bottom left: polarized intensity at $\nu$ = 950~MHz. Bottom right: fractional polarization at $\nu$ = 950~MHz. Fractional polarization is computed in pixels with $I_{\nu} \geq 0.5$\% of the peak at each frequency. Jet intensity hot spots near the head again reveal recollimation structures mentioned in the text. The $^*$ in labels serves as a reminder that the nominal length (and time) scales can be increased or decreased by adjustment of the bending length, $\ell_b$.}
\label{t135_pol}
\end{figure}
Figure \ref{t135_pol} displays the polarization properties associated with the 150~MHz and 950~MHz intensities in figure \ref{t135_obs}. Faraday rotation is ignored. Polarized flux comes primarily from the brightest mid-tail plumes mentioned in the previous paragraph and from recollimation features in the jets. Coherent jet emissions in the head region are highly polarized (mostly $\gtrsim70\%$) at both frequencies, because of  well-organized magnetic fields. Those field align with the jets (become poloidal) away from the launch cylinder due to stretching processes described in \S \ref{MagDyn}. Downwind of the disruption zone fractional polarization becomes patchy, and drops significantly in most areas. In the brightest regions discussed in the previous paragraph, fractional polarization is reduced even where the associated jets remain coherent. The jet flows are bent and twisted in this region, so most lines of sight pass through multiple, uncorrelated magnetic field alignments. The highest degrees of polarization in the tails are mostly on their edges, and, perhaps counter-intuitively, in fainter areas with short paths through emitting plasma. With good sensitivity and high spatial resolution, on the other hand, individual strongly magnetized and highly polarized filaments within the tails should still be observable.

\subsection{Evolution of Integrated Synthetic Radio Emissions}
\label{sec:specevolv}

\begin{figure}[ht]
    \centering
    \includegraphics[width=0.7\textwidth]{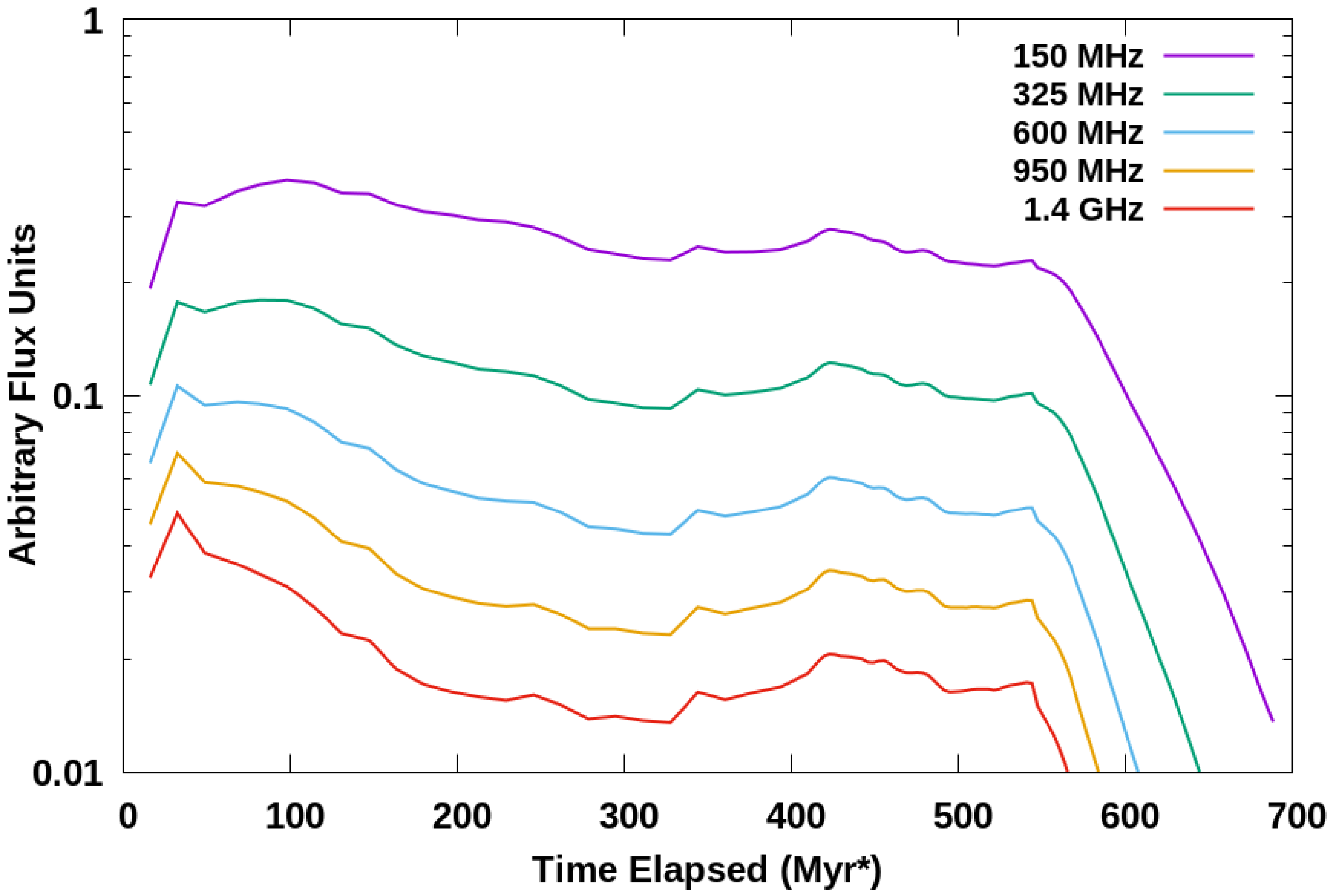}
    \caption{Time evolution of integrated fluxes (arbitrary units) at 150~MHz, 325~MHz, 600~MHz, 950~MHz and 1.4~GHz. Fluxes are sampled every 16.4~Myr$^*$ before 393.5~Myr$^*$, then every 3.28~Myr$^*$ thereafter. Jets turn off at $t = 546.6$~Myr$^*$. The~$^*$ in labels serves as a reminder that the nominal length (and time) scales can be increased or decreased by adjustment of the bending length, $\ell_b$.}
    \label{flxevo}
\end{figure}
\begin{figure}[ht]
    \centering
    \includegraphics[width=0.7\textwidth]{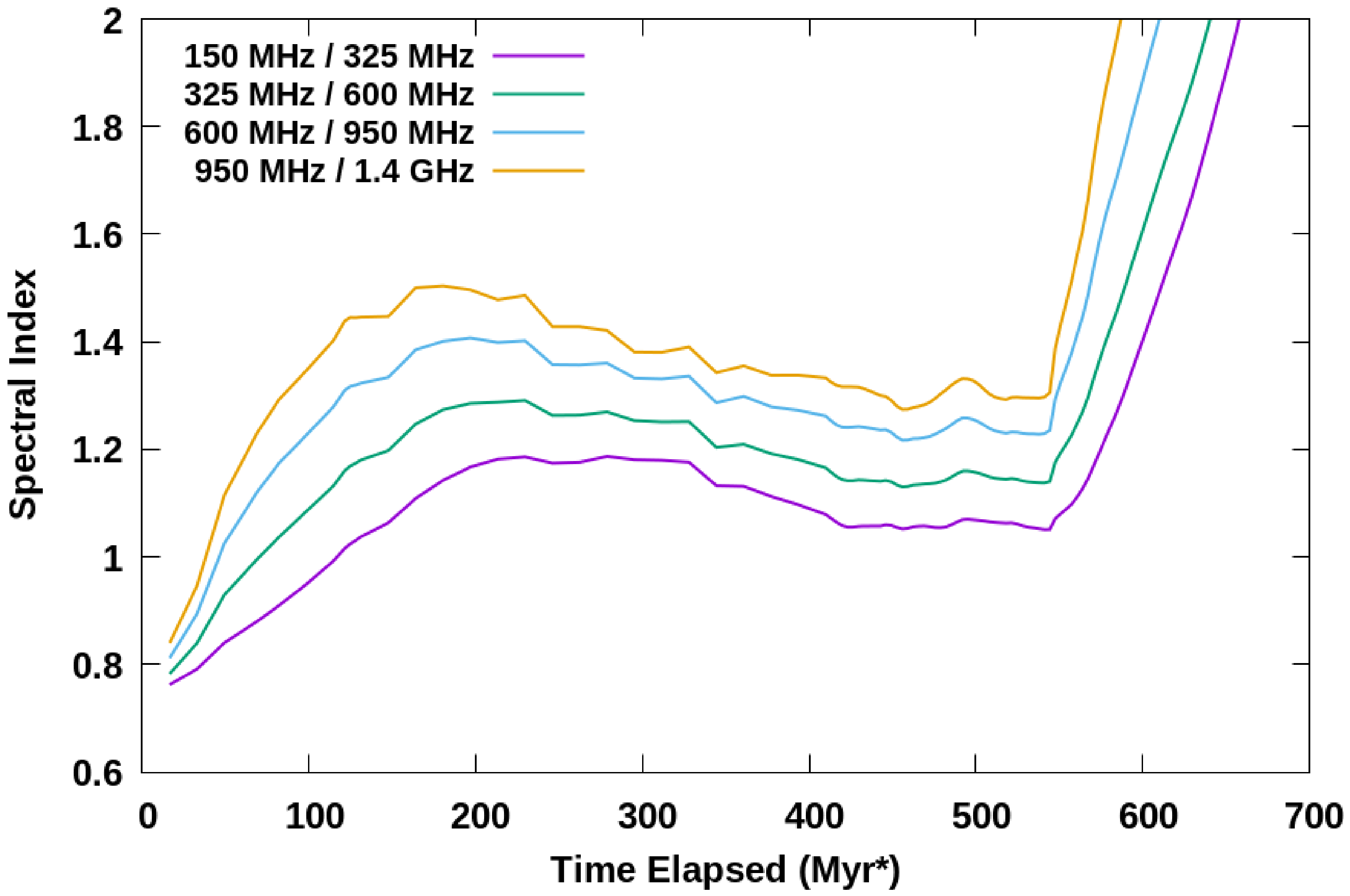}
	\caption{Time evolution of integrated spectral indices, $\alpha_{\nu 1,\nu 2}$, between adjacent frequencies in figure \ref{flxevo}, sampled at the same time interval. The $^*$  serves as a reminder that the nominal time scale can be increased or decreased by adjustment of the bending length, $\ell_b$.}
    \label{slopevo}
\end{figure}

Figures \ref{flxevo} and \ref{slopevo} display the time history of integrated synthetic radio emissions from our reference NAT. Figure \ref{flxevo} shows the evolution of source-integrated fluxes at 5 frequencies (150~MHz, 325~MHz, 600~MHz, 950~MHz and 1.4~GHz), while figure \ref{slopevo} illustrates the evolution of source-integrated spectral indices between pairs of these frequencies. The peak fluxes at the 5 frequencies are not simultaneous, but all come relatively early, before the NAT phase is established. There are obvious secondary and roughly coincident flux peaks at all 5 frequencies around $t \sim 420$~Myr, or shortly before the state represented in figure \ref{t135_obs}, i.e.,~coincident with a major jet disruption event. Perhaps the most notable property apparent in the integrated flux history is that following the initial early peaks, the fluxes at all 5 frequencies are relatively steady until jet activity ceases, $t \approx 550$~Myr. After that the fluxes drop roughly exponentially on a timescale $\sim 100$~Myr, consistent with our estimate for the iC cooling time, $\tau_{rad}$. The nearly steady fluxes over something close to 400 Myr is remarkable, and indicates a rough balance of radiative growth and decay factors. 

\begin{figure}[ht]
    \centering
    \includegraphics[width=0.5\textwidth]{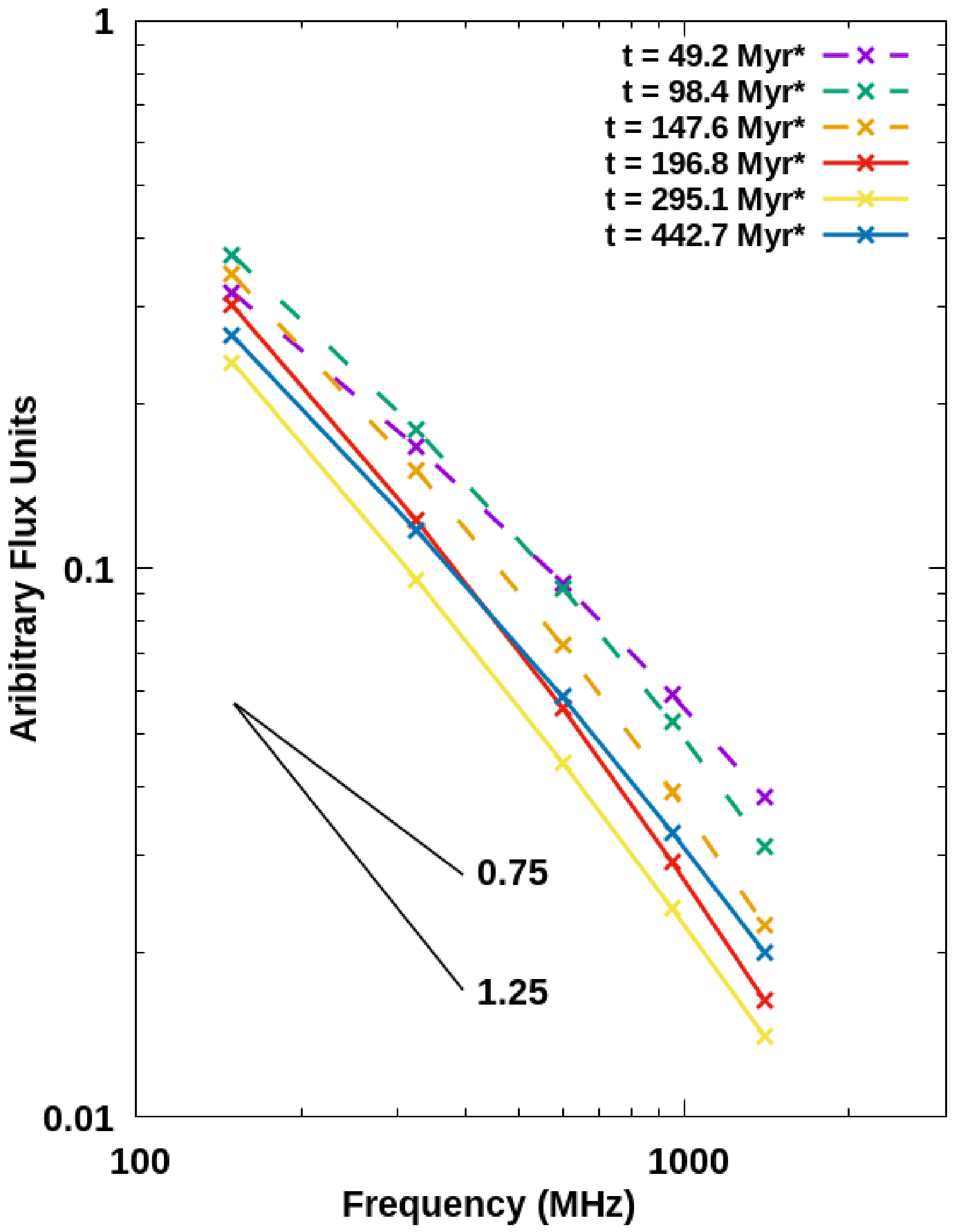}
	\caption{Integrated spectra from synthetic observations at selected times during simulation. The source population spectral slope, $\alpha_0$, and the fiducial $\alpha_0 + 0.5$ integrated spectral slope discussed in the text are shown for comparison. The $^*$ in labels serves as a reminder that the nominal time scale can be increased or decreased by adjustment of the bending length, $\ell_b$. }
    \label{IndivSpectra}
\end{figure}

That the fluxes at all 5 frequencies are approximately steady over a long time suggests, as well, that the integrated source spectra are not evolving dramatically during that time. Figures \ref{slopevo} and \ref{IndivSpectra} allow a closer look at this issue. Over the initial $\sim 200$ Myr of source evolution, so from startup through the transition into initial development of NAT morphology, there is significant steepening of the spectrum, increasing towards the higher frequencies; that is, the integrated spectrum ages during initial source development. The timescale for this is, not surprisingly, again comparable to the iC radiative cooling time. By $t \sim 200$ Myr, the spectral indices range between $\alpha_{150,325} \sim 1.2$ and $\alpha_{950,1400} \sim 1.5$. Subsequently, the slopes very slowly flatten slightly over the next $\sim 300$~Myr by amounts that are around $\sim 10$\%. Roughly speaking, the spectral shape of the source is, effectively self-similar from the time the NAT form develops until the jet activity ceases.  This behavior is also apparent in figure \ref{IndivSpectra}, which presents the integrated source synchrotron spectra between 150~MHz and 1.4~GHz at 6 times between 49.2~Myr and 442.7~Myr.

A closer look at the events reveals the following relevant information. During the initial phase of the source evolution, the spectrum steepens from its injection form ($\alpha = 0.75$)  mostly due to iC cooling of CRe in the initial, transient plumes, which dominate the emission during the first $\sim 200$~Myr. The subsequent spectral flattening comes as the initial plumes fade, and the structure transitions to an approximately steady-state NAT morphology for the duration of time the jet is active. After the jet activity ends, of course, the spectrum ages as CRe radiate away their remaining energy, most rapidly at high energies.

Additional insights to the spectral evolution of our source while it is a NAT driven by steady jets come from the classic treatment of synchrotron emissions from a CRe population that ages radiatively at a steady rate while maintaining an isotropic distribution and being refreshed at a steady rate. In particular, if an isotropic CRe population is injected with a constant power law momentum spectrum and uniform magnetic fields yielding a synchrotron emissivity spectral index $\alpha_0$, then above a critical frequency corresponding to CRe with radiative cooling time, $\tau_{rad} \approx t$, the integrated spectral index becomes $\alpha_0 + 0.5$, while below that frequency, the integrated spectral index remains the injection index, $\alpha_0$ \citep[e.g.,][]{pac70}. In this simple situation we would expect over a time span of several hundred Myr that our synthetic spectrum would steepen from $\alpha_0 = 0.75$ to a steeper power law with $\alpha = 1.25$ above some frequency decreasing over time. Interestingly, by the beginning of the steady phase of our NAT evolution, some 200~Myr in, $\alpha_{150,325} \sim 1.2$, then actually flattening slightly to $\alpha_{150,325} \sim 1.1$ by $t \sim 400$~Myr. These slope changes seem roughly in accord with the simple model just mentioned. On the other hand, the integrated spectrum is actually somewhat convex, becoming steeper at higher frequencies. That is not in accord with the simple model, at least after the low frequency spectrum has become ``aged.'' 

To put that in context it is important to keep in mind that the spectrum of the simple aging model just outlined assumes not only that the rate of aging and injection are constant, which is approximately true in our NAT, but also that the emissions come from a region of uniform magnetic field. This is certainly not true in our NAT, and leads to spatial variations in the spectra, as discussed above. Indeed, during the NAT phase the emissions from weak field regions, such as large portions of the tails, have distinctly steeper spectra because the CRe there are not being refreshed. In particular they have spectra steeper than $\alpha \sim 1.5$ at the higher frequencies. Although the surface brightness in those regions is relatively low, they still can contribute a significant portion of the integrated fluxes. The prolonged existence of a steady, self-similar, curved spectrum, on the other hand, runs counter to use of spectral shape as a direct age metric.

\section{Conclusion}

We have reported on an investigation of a simulated RG undergoing a long-term interaction with a persistent, homogeneous ICM wind perpendicular to the axis of the source AGN jets. The RG evolution ultimately gives rise to a narrow-angle head-tail (NAT) morphology. Such objects have been observed by radio astronomers for decades, and there have been multiple, prior theoretical and computational model efforts.  The present paper represents the first high-resolution numerical study focused on the detailed evolution of a NAT utilizing 3D MHD, including energy-dependent transport of relativistic CRe released by the AGN. This treatment allows us not only to follow detailed dynamics and topology of magnetic fields developed by the NAT coming from the AGN, but also to generate meaningful synthetic synchrotron observations allowing for energy evolution of the CRe. In the simulation studied here the ICM was unmagnetized in order to focus on the evolution of the AGN injected magnetic field. We also include in our report an Appendix exploring simple analytic treatments of the development of NAT morphology for a variety of wind-jet orientations. The key findings from this work are as follows.

\begin{enumerate}
\item The RG takes quite a long time, $\sim100$~Myr for this RG, to develop a NAT morphology. During the early phase, the jets grow transverse to the wind faster than the bending action by the wind ram pressure, and jet plasma is propelled to a transverse extent roughly twice the eventual equilibrium jet trajectory. This phase includes a transient stage where the jets are gently bent and the morphology resembles that of a WAT.
\item Post-bending, the NAT plasma supplied by the AGN jets, including CRe and magnetic fields, is not simply advected passively downstream. Rather, the deflected jets actively push downwind and power the growth of the tails, the ends of which extend faster than the wind speed so long as the jets remain active. For the geometry of this simulation all jet momentum pointing downstream has been transferred to the light jets from the heavier wind; momentum of the denser, slower wind has been concentrated into the lighter, faster jets. This momentum is then transferred to the mixed, intermediate density tails, leading to a net velocity boost of that flow over the undisturbed wind flow. This boost of the tail propagation could, in some situations lead to interactions between the AGN and regions well downwind of where interactions with the wind began, such as a pre-existing ICM shear layer or shock. This could include a shock that created the wind.
\item The jets themselves remain remarkably coherent once bent, although instabilities do arise that cause the jets to flap and sometimes disrupt on characteristic timescales determined by signal propagation between the tails, $\sim 25-50$~Myr in the NAT simulated here. These events are in our simulations mostly initiated by finite amplitude disturbances external to each jet, including chaotic motions in the jet wakes, and also from earlier disturbances propagated from flapping motions of the other jet. We speculate that this dynamical feedback between the two jets could be an essential element of NAT evolution. 
\item The dynamical interaction between the jets and the wind greatly enhances magnetic fields coming from the AGN in the NAT, primarily via stretching behavior. The plasma emerging from our simulated AGN has a toroidal magnetic field that is initially dynamically insignificant (high plasma $\beta_p$). However, stresses coming from the bending process quickly give rise to a mostly poloidal magnetic field with the strongest fields having $\beta_p \lesssim 1$. The jet disruption events further stretch the fields and increase magnetization of the deposited plasma. As a result, the strongest magnetic fields in the tails are found in magnetic filaments that in the simulations are comparable to the numerical resolution  thick (so, $\sim$ kpc) and tens of kpc long. 
\item Observationally, once CRe populations released early on have significantly aged via radiative losses, the synthetic synchrotron radiation produced by our NAT exhibits self-similar, steady-state spectral properties for 100's of Myr, until the AGN activity ceases. The integrated spectrum during this period appears slightly convex and remarkably steady. This combination of a steady, but curved form means the curvature does not provide a useful, direct age metric of the source.
\end{enumerate}

This research was supported at the University of Minnesota by NSF grant AST1714205 and through resources provided by the Minnesota Supercomputing Institute. CN was supported by an NSF Graduate Fellowship under Grant 00039202. Thanks to our many colleagues, particularly Larry Rudnick and Avery F. Garon for their motivation and insight. We also thank an anonymous referee for a careful reading of the manuscript and useful comments that helped us  improve the presentation of our work.

\appendix

\section{Modeling NAT Head Jet Trajectories}

Here we outline and compare with simulations  a simple, but flexible formalism for modeling the approximate, steady-state trajectories near their source of supersonic jets propagating in a crosswind for an arbitrary orientation between the jet flow at its source and the wind flow. The formalism is consistent with the classic jet bending treatment for orthogonal crosswinds \cite{brb79,jnom17}, but, because jet-wind alignments can take any orientation, and because 3D simulations reveal rather complex flow dynamics near the jets in this situation, the formalism is generalized and designed for easy empirical tuning. For simplicity, the jet velocity is assumed to have constant magnitude, $v_j$, while the direction of the jet flow, $\hat{v}_j$, is deflected by pressure gradients across the jet set up by the wind flow.

We apply a Cartesian coordinate system similar to that in the main text of this paper, in which the upwind, unperturbed wind has velocity $\vec{v}_w = v_w \hat{x}$, with an unperturbed wind density, $\rho_w$. We express the interactions in terms of a uniform wind and steady jet sources, but because the model is ``local'' it is, in principle, straightforward to extend to nonuniform winds and time varying jet sources. We assume the jets emerge from their source with velocity, $\vec{v}_j = v_j(\cos{\theta_{ji}}\hat{x} + \sin{\theta_{ji}}\hat{z})$, with $i = 1,2$ to account for two jets and with $0\le \theta_{ji}< 2\pi$. Thus, for $\theta_{ji} = 0 ~(\pi)$ an emergent jet and the wind are aligned and parallel (anti-parallel), while for $\theta_{ji} = \pi/2,~3\pi/2$, the jet and wind flows are orthogonal. We assume for simplicity that the jet density, $\rho_j$, and jet radius, $r_j$, are constant along the jet; only the direction, $\theta$, of $\hat{v}_j$ varies as the jets propagate.  Except at its source we hereafter drop the subscript $j$ from the jet propagation direction, $\theta$. That is, $\theta_{ji}$ always refers to a jet as it is launched. The first index, $j$, indicates a launching jet, while the second index, i = 1,2, discriminates between the two jets.  In our model and its test simulations we include jet pairs satisfying $\theta_{j2} = \theta_{j1} + \pi$.

The incidence of a crosswind on a jet leads to a pressure enhancement on the upwind side of the jet relative to the downwind side. This pressure drop across the jet deflects the jet trajectory towards the downwind direction. For consistency with our assumption that the jet speed, $v_j$, is constant, we assume in our simple model that the pressure gradient induced by the crosswind is perpendicular to the local jet axis and in the jet-wind plane; that is in the direction $\hat{a}_{\perp} =\pm ( \sin{\theta}\hat{x} -\cos{\theta}\hat{z})$, where the upper (lower) sign applies for $0<\theta<\pi$ ($\pi < \theta < 2\pi$). Even though wind density and pressure distributions near the jets in our simulations are neither simple nor really steady, we still expect on average that the magnitude of the effective pressure gradient will scale approximately with the wind ram pressure, $\rho_w v_w^2$, and inversely with the jet radius, $r_j$, with allowance for  influence of projection through $\theta$. We combine these considerations to produce an estimate for the effective pressure gradient across the jet, $\alpha \rho_w v_w^2 f(\theta)/r_j$, where $\alpha \sim 1$ is a parameter, perhaps somewhat dependent on the specific situation, while $f(\theta)$ is a function we will define below intended to account for local jet orientation with respect to the wind flow. Then the acceleration of jet plasma normal to its propagation is $|a_{\perp}| \approx \alpha \rho_w v_w^2 f(\theta)/(r_j\rho_j)$.

Using the above relations we can estimate an effective local bending radius for the jet, $R \approx ~v_j^2/|a_{\perp}|$ $ \approx~\ell_b / (\alpha f(\theta))$  where we have defined $\ell_b$ as  a characteristic ``{\emph{bending length}}'',

\begin{equation}
\ell_b = r_j  \frac{\rho_j v_j^2}{\rho_w v_w^2}.
\end{equation}

We prefer the term ``bending length'' as a somewhat more appropriate concept in describing our model jet trajectories than ``bending radius,'' since the trajectories are not generally circular arcs.
Since, however, the effective acceleration is assumed normal to the local jet flow, we can use these results to establish the relation for the rate in time at which the local jet flow is deflected,

\begin{equation}
\frac{d \theta}{d t} = \pm \frac{\alpha v_j}{R} = \pm \frac{ v_j}{\ell_b} f(\theta).
\label{eq:bendrate}
\end{equation}

The upper (lower) sign applies when the deflection is clockwise (counter clockwise). 
Since we generally start with pairs of oppositely directed jets, $\theta_{j2} = \theta_{j1} + \pi$, unless $\theta_{j1} = 0$, we  have one jet bending clockwise ($d \theta/dt > 0$) and one bending counter clockwise ($d \theta/dt < 0$).
 Since $dx/dt = - v_j \cos{\theta}$ and $dz/dt = v_j \sin{\theta}$, we can use equation \ref{eq:bendrate} to determine the trajectory of the jet in terms of $\theta$ as

\begin{eqnarray}
\label{eq:delx}
x^{\prime} - x_j^{\prime} =\frac{x-x_j}{\ell_b} =  -  \frac{1}{\alpha} \int_{\theta_j}^{\theta} \frac{\cos{\theta}}{f(\theta)} d\theta =- \frac{1}{\alpha} \int_{\sin{\theta_j}}^{\sin{\theta}} \frac{d \sin{\theta}}{f(\theta)} = - \frac{1}{\alpha} I_x(\theta,\theta_j) ,\\
\label{eq:dely}
z^{\prime} - z_j^{\prime} = \frac{y - y_j}{\ell_b} =  \frac{1}{\alpha} \int_{\theta_j}^{\theta}\frac{\sin{\theta}}{f(\theta)} d\theta = \frac{1}{\alpha} I_z(\theta,\theta_j),
\end{eqnarray}

where $(x_j,z_j)$ represents the jet source coordinates, and we have introduced the bending length normalized coordinates, $x^{\prime} = x/\ell_b,~ z^{\prime}= y/\ell_b$, while $I_x$ and $I_z$ encode the dependence of the model on $f(\theta)$. For the trivial, but unphysical case, $f(\theta) = 1$, $I_x = \sin{\theta} - \sin{\theta_j}$, $I_z = \cos{\theta_j} - \cos{\theta}$, and the trajectory becomes a circle of radius, $R^{\prime} =  R/\ell_b =1/\alpha$, as one would expect. In interpreting equations \ref{eq:delx} and \ref{eq:dely} we should keep in mind that according to the conditions outlined above $\theta -\theta_j$ can be positive or negative.

Physically, we would expect the induced normal force to increase as the jet and wind flows become more nearly orthogonal; that is, as $|\hat{v}_j\times \hat{v}_w| = |\sin{\theta}|$ increases.  Although we are not aware of any generally applicable physical form for $f(\theta)$, the simple form $f(\theta) =| \sin{\theta}^n|$  with $n = 1,2,3,...$ provides a convenient, physically consistent form that is also easy to integrate to obtain $I_x$ and $I_z$.  It is intuitively obvious that smaller values for $n$ lead to more sharply bent jets than larger values for $n$. Quantitatively the forms become, for $n = 1$, $I_x = \ln{(\sin{\theta}/\sin{\theta_{ji}})}$, $I_z = (\theta - \theta_{ji})$, for $n = 2$, $I_x =( 1/\sin{\theta_{ji}} - 1/\sin{\theta})$, $I_z = \ln{(\tan{\theta/2}/\tan{\theta_{ji}/2})}$, and for $n = 3$, $I_x = (1/2)\left[1/(\sin{\theta_{ji}})^2  - 1/(\sin{\theta})^2\right]$, $I_z = (\cot{\theta_{ji}} - \cot{\theta})$.

\begin{figure}[!ht]
\includegraphics[width=\textwidth]{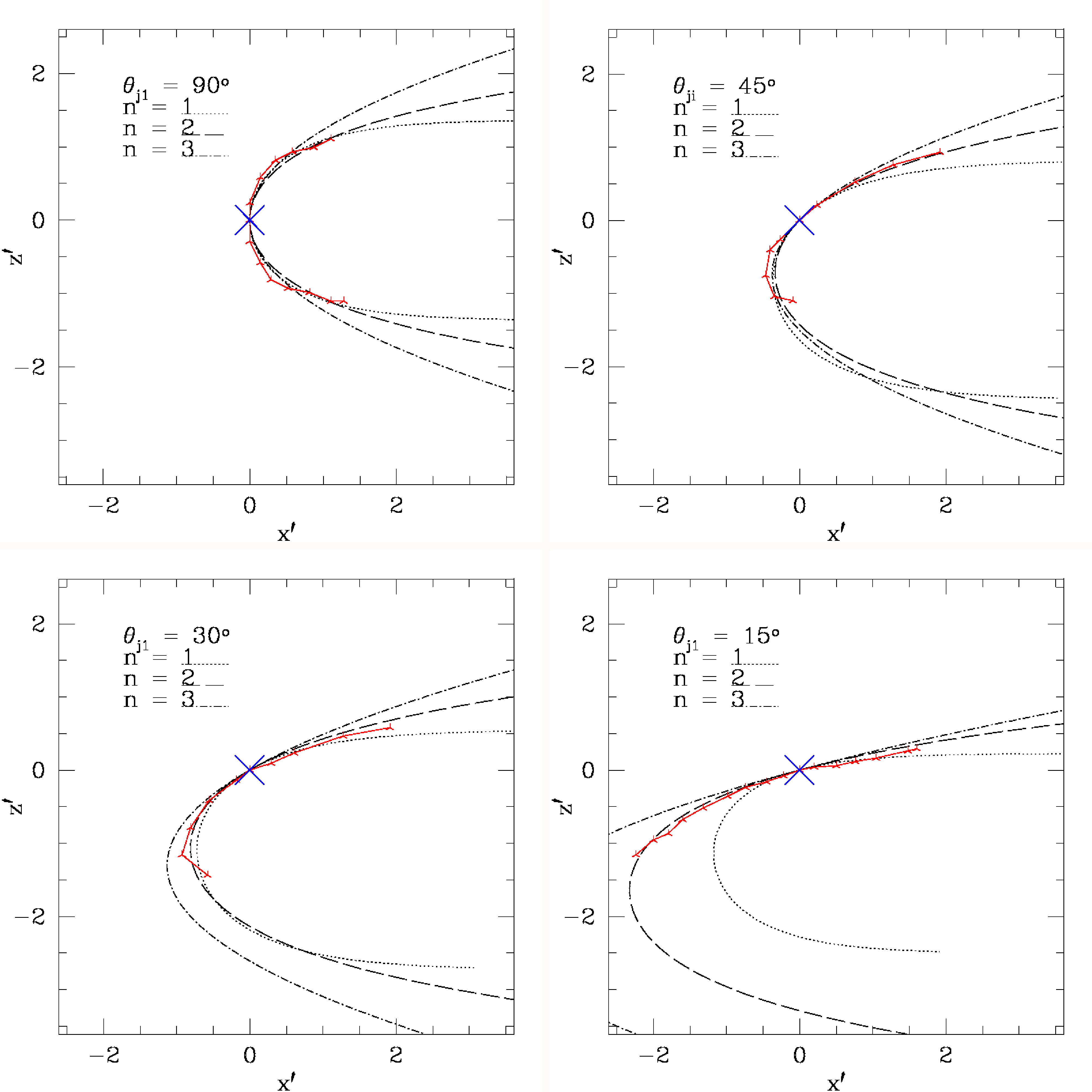}
\caption{Shapes in normalized coordinates, $x^{\prime}$, $z^{\prime}$ computed using equations \ref{eq:delx} and \ref{eq:dely} with $f(\theta) = |\sin{\theta}^n|$ with $n = 1, 2$ and $3$ (black curves). The ram pressure efficiency factor, $1 \la \alpha\la 1.3$, depending on both $\theta_{ji}$ and $n$. The jet source is marked with an `X'.  Red points and curves trajectories measured from 3D MHD simulations similar to those described in \S \ref{SimDet} over portions of the ``head region''empirically  deemed  quasi-steady after jets had propagated beyond the head  into the tails.}
\label{fig:natplots}
\end{figure}

\begin{figure}[!ht]
\includegraphics[width=\textwidth]{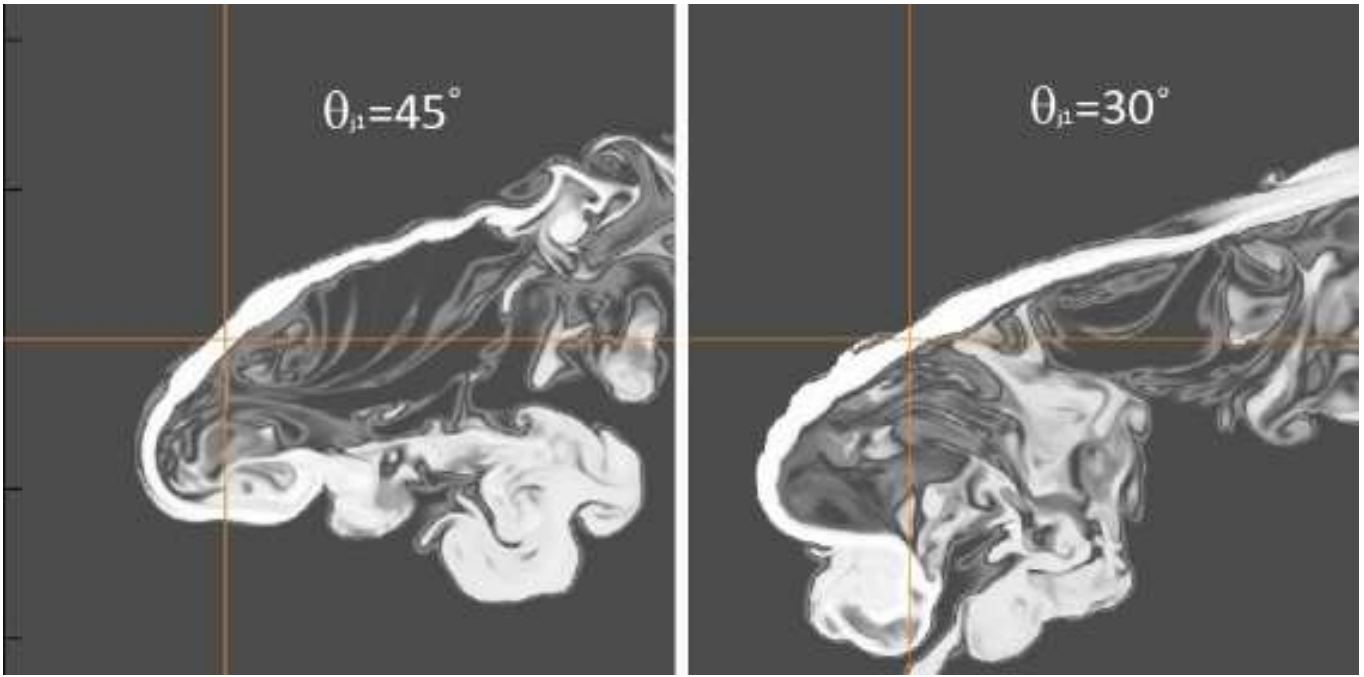}
\caption{2D (x-z plane) slices showing $C_j$ in two 3D MHD simulations used in comparing the model trajectories shown in Figure \ref{fig:natplots}. Cross hairs identify the jet source center.  Simulation jet trajectory coordinates shown in Figure \ref{fig:natplots} were measured from the ends of the jet launch cylinder to the center of the propagating jet. }
\label{fig:natpair}
\end{figure}

Figure \ref{fig:natplots} illustrates the solutions to equations \ref{eq:delx} and \ref{eq:dely} for $\theta_{j1} = 90^{\degree}, 45^{\degree}, 30^{\degree}$ and $15^{\degree}$ with $n = 1, 2, 3$. The red curve and data points in each panel show the jet trajectories from 3D MHD test simulations for each $\theta_{j1}$, $\theta_{j2}$ in the domains where they remained steady in time after the jet termini extended into the tail regions. As discussed in \S \ref{evol-outline},  when jets in our simulations became strongly bent, they developed intermittent ``flapping'' behaviors, even though they typically remained coherent well into the tails. For each jet orientation and $n$  the $\alpha$ parameter in the plotted curve in Figure  \ref{fig:natplots} was set to yield the best empirical fit to simulation data with the same $\ell_b$ and $\theta_{j1}$. In general $1\la\alpha\la 1.3$, with $\alpha$ increasing slowly  from $\theta_{j1} = \pm \pi/2$ as $\theta_{j1} \rightarrow 0$.  For validation, figure \ref{fig:natpair} shows snapshot 2D views from simulated flows used to trace jet trajectories in the $\theta_{j1} = 45^{\degree}$ and $\theta_{j1} = 30^{\degree}$ orientations shown in figure \ref{fig:natplots}. Analogous simulation behaviors for $\theta_{j1} = 90^{\degree}$ are available, for example, in figure \ref{wagfig}  and for $\theta_{j1} = 15^{\degree}$ in \citep[][]{nolt19}. 
Although not presented here, we mention that we have conducted other simulations similar to those cited using a wide range of individual wind and jet Mach numbers. General behaviors did not depend noticeably on the individual Mach numbers, but only on their ratio. Similarly, \cite{nolt19} demonstrated for shock interactions with AGN jets, that, so long as the jets are sufficiently supersonic, the resulting dynamics can be predicted reliably using the Mach number ratio without specifying the individual Mach numbers.

We conclude that these simple jet trajectory models provide reasonable matches to our simulation results over wide a range of jet orientations, with parameters $n \sim 2$ and $\alpha \ga 1$, but caution that past the point where a jet has been deflected by $\Delta\theta \ga \pi/2$ jet trajectories tend to be quite unsteady. Even then, the time average behaviors match the models reasonably well.Beyond what we show here, we have conducted multiple simulation tests in which we varied individual parameters, including jet density, $\rho_j$ and velocity, $v_j$, as well as wind density, $\rho_w$ and velocity, $v_w$. The steady jet trajectories cited here were robust, so long as they were expressed in terms of the normalized scales, $x^{\prime}$ and $z^{\prime}$ defined in this Appendix.

\begin{thebibliography}{99}
\bibitem[Balsara \& Norman(1992)]{bal92}Balsara, D. \& Norman, M. L., 1992, \apj, 393, 631
\bibitem[Begelman et al.(1979)]{brb79}Begelman, M. C., Rees, M. J. \& Blandford, R. D., 1979, Nature, 279, 770
\bibitem[Bicknell(1994)]{bicknell94}Bicknell, G. V., 1994, \apj, 422, 542
\bibitem[Bodo \& Tavecchio (2018)]{bodo18}Bodo, G \& Tavecchio, F. 2018, A\&A, 609, A122
\bibitem[Bonafede et al.(2014)]{bona14}Bonafede, A., Intema, H T., Brueggen, M., Girardi, M., Nonino, M., Kantharia, N., van Weeren, R. J. \& Roettgering, H. J. A., 2014, \apj, 785, 1
\bibitem[Brunetti \& Jones(2014)]{bj14}Brunetti, G. \& Jones, T. W., 2015, IJMPD, 23, 1430007-98
\bibitem[Brunetti \& Lazarian(2011)]{brunetti11}Brunetti, G. \& Lazarian, A., 2011, \mnras, 412, 817
\bibitem[Burns et al.(1979)]{burns79}Burns, J., Owen, F., Rudnick, L. \& Greisen, E., 1979, \apj, 229, 590
\bibitem[Gan et al.(2017)]{gan17}Gan, Z., Li, H., Li, S. \& Yuan, F., 2017, \apj, 839, 14
\bibitem[Garon et al.(2019)]{garon19}Garon, A., Rudnick, L., Wong, O. I., Jones, T. W., Kim, J., Andernach, H., Shabala, S., Kapinska, A. D., Norris, R. P., de Gasperin, F., Tate, J. \&  Tang, H., 2019, \aj, 157, 17
\bibitem[Heinz et al.(2006)]{heinz06}Heinz, S., Brueggen, M., Young, A. \& Levesque, E. 2006, \mnras, 373, 65
\bibitem[Hoang et al.(2017)]{hoang17}Hoang, D. N., Shimwell, T. W., Stroe, A., et al, 2017, \mnras, 471, 1107
\bibitem[Jones \& Kang(2005)]{jk05}Jones, T. W. \& Kang, H., 2005, Astropart. Phys., 24, 75
\bibitem[Jones et al(2017)]{jnom17}Jones, T. W., Nolting, C., O'Neill, B. J. \& Mendygral, P. J. 2017, Physics of Plasmas, 24, 41402
\bibitem[Jones \& Owen(1979)]{jo79}Jones, T. W. \& Owen, F. N., 1979, \apj, 234, 818
\bibitem[Kang \& Ryu(2016)]{kr16}Kang, H. \& Ryu, D., 2016, \apj, 823, 13
\bibitem[Laing \& Bridle(2008)]{laing08}Laing, R. \& Bridle, A., 2008, ``Extragalactic Jets: Theory and Observation from Radio to Gamma Ray,'' (ASP Conference Series, Vol 386) (T. Rector \& D. De Young, Eds.) p 70
\bibitem[Landau \& Lifshitz (1987)]{landl} Landau, L. and Lifshitz, E., 1987, ``Fluid Mechanics,'' (Pergamon Press, Oxford), ch 1.
\bibitem[Loken et al.(1995)]{loken95}Loken, C., Roettiger, K., Burns, J. \& Norman, M. 1995, \apj, 445, 80
\bibitem[Lynn et al.(2014)]{lynn14}Lynn, W. L., Quataert, E., Chandran, B. \& Parrish, I., 2014, \apj, 791, 71
\bibitem[Mendygral(2011)]{mendthesis}Mendygral, Peter John. (2011). PhD thesis. Simulations and Synthetic Observations of Active Galactic Nuclei Jets in Galaxy Clusters: Numerical Tools and Experiments.. Retrieved from the University of Minnesota Digital Conservancy, http://hdl.handle.net/11299/113253
\bibitem[Morsony et al.(2013)]{mors13}Morsony, B., Miller, J. J., Heinz, S., Freeland, E., Wilcots, E. Brueggen, M. \& Ruszkowski, M. 2013, \mnras, 431, 781
\bibitem[Miniati et al.(2001)]{min01}Miniati, F., Jones, T. W., Kang, H. \& Ryu, D., 2001, \apj, 562, 233
\bibitem[Morsony et al.(2013)]{morsony13}Morsony, B. J., Miller, J. J., Heinz, S., Freeland, E., Wilcotts, E., Bruggen, M., Ruszkowski, M., 2013, \mnras, 431, 781
\bibitem[Nolting et al.(2019a)]{nolt19}Nolting, C., Jones T. W., O'Neill, B. J. \& Mendygral, P. J., 2019, \apj, 876, 154
\bibitem[Nolting et al.(2019b)]{nolt19b}Nolting, C., Jones, T. W., O'Neill, B. J. \& Mendygral, P. J., 2019 \apj (in preparation)
\bibitem[Norman et al.(1982)]{norman82}Norman, M. L., Smarr, L., Winkler, K-H. \& Smith, M. D., 1982, A\&A, 113, 285
\bibitem[O'Neill et al.(2019)]{on19b}O'Neill, B. J., Jones, T. W., Nolting, C. \& Mendygral. P. J., 2019 \apj (in preparation)
\bibitem[Owen et al.(2014)]{owen14}Owen, F. N., Rudnick, L., Eilek, J., Rau, U., Bhatnagar, S. \& Kogan, L., 2014, \apj, 794, 24
\bibitem[Pacholczyk(1970)]{pac70}Pacholczyk, A. G. 1970, Radio astrophysics: Nonthermal Processes in Galactic and Extragalactic Sources
\bibitem[Perucho(2013)]{perucho13}Perucho, M., 2013, arXiv:1308.6168v2
\bibitem[Porter et al.(2009)]{porter09}Porter, D. H., Mendygral, P. J. \& Jones, T. W., 2009, AIPC, 1201,259
\bibitem[Rudnick \& Owen(1977)]{rudnick77}Rudnick, L. \& Owen, F. N., 1977, \aj, 82, 1
\bibitem[Rybicki \& Lightman(1979)]{rybicki79}Rybicki, G. \& Lightman, A., 1979, ``Radiative Processes in Astrophysics,'' Ch 6 (John Wiley \& Sons) (New York)
\bibitem[Ryu \& Jones(1995)]{rj95}Ryu, D., \& Jones, T. W., 1995, \apj, 442, 228
\bibitem[Ryu et al(1998)]{ryu98}Ryu, D., Miniati, F., Jones, T. W. \& Frank, A., 1998, \apj, 509, 244
\bibitem[Sarazin(1999)]{szin99}Sarazin, C. L., 1999, \apj, 520, 529S
\bibitem[Shimwell et al.(2015)]{shim15}Shimwell, T. W., Markevitch, M., Brown, S., Farrar, G. R., Feretti, L., Gaensler, B. M., Johnston-Hollitt, M., Lage, C. \& Srinivasan, R., 2015, \mnras, 449, 1486
\bibitem[van Weeren et al.(2017)]{vanw17}van Weeren, R. J., Andrade-Santos, F., Dawson, W. A., Golovich, N., Lal, D.V., Kang, H., Ryu, D., Brueggen, M., Ogrean, G. A., Forman, W. R., Jones, C., Placco, V. M., Santucci, R. M., Wittman, D., Jee, M. J., Kraft, R. P., Sobral, D., Stroe, A. \& Fogarty, K., 2017, Nature Astronomy, 1,  5
\bibitem[Wilber et al.(2018)]{wilb18}Wilber, A., Br\"{u}ggen, M., Bonafede, A., Savini, F., Shimwell, T., van Weeren, R. J., Rafferty, D., Mechev, A. P., Intema, H., Andrade-Santos, F., Clarke, A. O., Mahony, E. K., Morganti, R., Prandoni, I., Brunetti, G., R\"{o}ttgering, H., Mandal, S., de Gasperin, F. \& Hoeft, M., \mnras, 473, 3536
\bibitem[Williams \& Gull(1984)]{wg84}Williams, A. G. \& Gull, S. F., 1984, Nature, 310, 33
\end{thebibliography}
\end{document}